\documentclass[a4paper,12pt]{article}

\usepackage{cite}
\usepackage{epsfig}
\usepackage{amssymb,amsmath}
\usepackage{mathrsfs}
\usepackage{bbm}
\usepackage{graphicx,subfigure}
\usepackage{amsmath}
\allowdisplaybreaks

\pdfoutput=1
\newlength{\dinwidth}
\newlength{\dinmargin}
\newcommand{\cR}{\mathcal{R}}

\newcommand{\Bqll}{B_{s,d}^0 \to \ell^+\ell^-}

\newcommand{\Bsmm}{B_s^0 \to \mu^+\mu^-}
\newcommand{\Bdmm}{B_d^0 \to \mu^+\mu^-}

\def\be{\begin{equation}}
\def\ee{\end{equation}}
\def\beqn{\begin{eqnarray}}
\def\eeqn{\end{eqnarray}}
\def\no{\nonumber}
\def\ba{\begin{array}{c}}
\def\bat{\begin{array}{cc}}
\def\ea{\end{array}}
\def\bi{\begin{itemize}}
\def\ei{\end{itemize}}

\def\cL{{\cal L}}

\def\cC{{\cal C}}

\newcommand{\eqn}[1]{(\ref{#1})}
\newcommand{\bel}[1]{\be\label{#1}}

\usepackage{color}

\definecolor{Brown}{rgb}{0.5,0.25,0}

\begin{document}

\title{
\begin{flushright}\vbox{\normalsize FTUV/14$-$0423 \\[-3pt] IFIC/14$-$05}
\end{flushright}\vskip 20pt
{\bf $\boldsymbol{\Bqll}$ Decays in the \\[10pt] Aligned Two-Higgs-Doublet Model}}
\bigskip
\author{Xin-Qiang Li$^{1,2}$$\footnote{xqli@itp.ac.cn}$, Jie Lu$^{3}$$\footnote{lu.jie@ific.uv.es}$ and Antonio Pich$^{3}$$\footnote{pich@ific.uv.es}$\\
{$^1$\small Institute of Particle Physics and Key Laboratory of Quark \& Lepton Physics~(MOE), }\\[-0.2cm]
{    \small Central China Normal University, Wuhan, Hubei 430079, P.~R.~China}\\[-0.1cm]
{$^2$\small State Key Laboratory of Theoretical Physics, Institute of Theoretical Physics,}\\[-0.2cm]
{    \small Chinese Academy of Sciences, Beijing 100190, China}\\[-0.1cm]
{$^3$\small IFIC, Universitat de Val\`encia -- CSIC, Apt. Correus 22085, E-46071 Val\`encia, Spain}}

\date{}
\maketitle
\bigskip \bigskip
\vspace{-0.5cm}

\begin{abstract}
{\noindent}The rare decays $\Bqll$ are analyzed within the general framework of the aligned two-Higgs doublet model. We present a complete one-loop calculation of the relevant short-distance Wilson coefficients, giving a detailed technical summary of our results and comparing them with previous calculations performed in particular limits or approximations. We investigate the impact of various model parameters on the branching ratios and study the phenomenological constraints imposed by present data.
\end{abstract}

\newpage

\section{Introduction}
\label{sec:intro}

The recent discovery of a Higgs-like boson~\cite{Aad:2012tfa,Chatrchyan:2012ufa}, with properties compatible with the Standard Model~(SM) expectations~\cite{Heinemeyer:2013tqa,Dawson:2013bba,Higgs:review}, is one of the greatest achievements in the past decades in particle physics and represents a major confirmation of our present theoretical paradigm. The LHC data suggest that the electroweak symmetry breaking~(EWSB) is probably realized in the most elegant and simple way, \textit{i.e.}, via the Higgs mechanism implemented through one scalar $\mathrm{SU}(2)_{\mathrm{L}}$ doublet. An obvious question we are now facing is whether the discovered $126~\mathrm{GeV}$ state corresponds to the unique Higgs boson incorporated in the SM, or it is just the first signal of a much richer scenario of EWSB. None of the fundamental principles of the SM forbids the possibility of an enlarged scalar sector associated with the EWSB.

Among the many possible scenarios for new physics~(NP) beyond the SM, the two-Higgs doublet model~(2HDM)~\cite{Lee:1973iz} provides a minimal extension of the scalar sector that naturally accommodates the electroweak~(EW) precision tests, giving rise at the same time to a large variety of interesting phenomenological effects~\cite{2HDM:review}. The scalar spectrum of the model consists of two charged fields, $H^\pm$, and three neutral ones, $h$, $H$ and $A$, one of which is to be identified with the Higgs-like boson found at the LHC. The direct search for these additional scalar states at high-energy collisions, or through indirect constraints via precision flavour experiments, is an important task for the next years. This will also be helpful to gain further insights into the scalar sector of supersymmetry~(SUSY) and other models with similar scalar contents.

Within the SM, flavour-changing neutral current~(FCNC) interactions are forbidden at tree level, and highly suppressed at higher orders, due to the Glashow--Iliopoulos--Maiani~(GIM) mechanism~\cite{GIM}. In a generic 2HDM, however, tree-level FCNC interactions generally exist, through non-diagonal couplings of neutral scalars to fermions. The unwanted FCNCs can be eliminated, imposing on the Lagrangian an ad-hoc discrete $\mathcal{Z}_2$ symmetry; depending on the different possible $\mathcal{Z}_2$ charge assignments, this results in four types of 2HDMs~(I, II, X and Y)~\cite{2HDM:review}, all satisfying the hypothesis of natural flavour conservation~(NFC)~\cite{Glashow:1976nt}. A more general alternative is to assume the alignment in flavour space of the Yukawa matrices for each type of right-handed fermions~\cite{Pich:2009sp}. The so-called aligned two-Higgs doublet model~(A2HDM) results in a very specific structure, with all fermion-scalar interactions being proportional to the corresponding fermion masses. It also contains as particular cases the different versions of the 2HDM with NFC, while at the same time introduces new sources of CP violation beyond the Cabibbo--Kobayashi--Maskawa~(CKM) phase~\cite{CKM}. These features make the A2HDM a very interesting theoretical framework, which leads to a rich and viable phenomenology, both in high-energy collider experiments~\cite{A2HDM:collider1,A2HDM:collider2}, as well as in low-energy flavour physics~\cite{Jung:2010ik,A2HDM:flavour}.

In the field of rare B-meson decays, the purely leptonic processes $\Bqll$, with $\ell=e$, $\mu$ or $\tau$, play an outstanding role in testing the SM and probing physics beyond it, because they are very sensitive to the mechanism of quark-flavour mixing. Within the SM, the FCNC transition is mediated by a one-loop amplitude, suffers from a helicity-suppression factor $m_{\ell}/m_b$, and is characterized by a purely leptonic final state. The first two features result in a double suppression mechanism, responsible for the extremely rare nature of these decays. The third feature implies that these processes are theoretically very clean, with the only hadronic uncertainty coming from the B-meson decay constants $f_{B_{s,d}}$. All these considerations make the rare leptonic decays $\Bqll$ a formidable probe of physics beyond the SM, especially of models with a non-standard Higgs sector like multi-Higgs doublet models~\cite{Logan:2000iv,Huang:2000sm,Bobeth:2001sq,Bqll:2HDM,Chankowski:2000ng} as well as various SUSY scenarios~\cite{Huang:2000sm,Bobeth:2001sq,Chankowski:2000ng,Dreiner:2012dh,Bqll:SUSY}.

As far as the experimental side is concerned, the decay modes with $\ell=\mu$ are especially interesting because the corresponding final state can be easily tagged. Over the last decade the upper bounds for the branching ratios of these decays have been improving continuously, thanks to the CDF and $\mathrm{D\O}$ collaborations at the Tevatron and, more recently, the ATLAS, CMS and LHCb experiments at the LHC~\cite{Albrecht:2012hp}. In November 2012, the LHCb experiment reported the first evidence of the decay $\Bsmm$, at the $3.5\,\sigma$ level~\cite{Aaij:2012nna}. The signal significance has been raised, respectively, to $4.0\,\sigma$ and $4.3\,\sigma$ by LHCb and CMS, after analyzing the currently available data set, with the averaged time-integrated branching ratio given by
\begin{align} \label{eq:Bsmmexp1}
 \overline{\mathcal{B}}(\Bsmm) &=
  \begin{cases}
   \left(2.9^{\, +1.1}_{\, -1.0}({\rm stat.})^{\, +0.3}_{\, -0.1}({\rm syst.})\right) \times 10^{-9} & \text{LHCb~\cite{Aaij:2013aka}}\\[0.2cm]
   \left(3.0^{\, +1.0}_{\, -0.9}\right) \times 10^{-9} & \text{CMS~\cite{Chatrchyan:2013bka}}
  \end{cases}\, ,
\end{align}
where the CMS uncertainty includes both the statistical and systematic components, but is dominated by the statistical uncertainties. The two measurements lead to the weighted world average~\cite{CMSandLHCbCollaborations:2013pla}
\begin{equation} \label{eq:Bsmmexp}
 \overline{\mathcal{B}}(\Bsmm)_{\rm exp.} = (2.9 \pm 0.7) \times 10^{-9}\,.
\end{equation}
At the same time, the branching fraction of $\Bdmm$ has also been determined with a signal significance of $2\,\sigma$ by the two experiments:
\begin{align} \label{eq:Bdmmexp1}
 \overline{\mathcal{B}}(\Bdmm) &=
  \begin{cases}
   \left(3.7^{\, +2.4}_{\, -2.1}({\rm stat.})^{\, +0.6}_{\, -0.4}({\rm syst.})\right) \times 10^{-10} & \text{LHCb~\cite{Aaij:2013aka}}\\[0.2cm]
   \left(3.5^{\, +2.1}_{\, -1.8}\right) \times 10^{-10} & \text{CMS~\cite{Chatrchyan:2013bka}}
  \end{cases}\, .
\end{align}
The corresponding combined result reads~\cite{CMSandLHCbCollaborations:2013pla}
\begin{equation} \label{eq:Bdmmexp}
 \overline{\mathcal{B}}(\Bdmm)_{\rm exp.} = \left(3.6^{\, +1.6}_{\, -1.4}\right) \times 10^{-10}\,.
\end{equation}
These measurements are in remarkable agreement with the latest updated predictions within the SM~\cite{Bobeth:2013uxa}:
\begin{equation} \label{eq:BqmmSM}
 \overline{\mathcal{B}}(\Bsmm) = (3.65 \pm 0.23) \times 10^{-9}\,, \qquad \overline{\mathcal{B}}(\Bdmm) = (1.06 \pm 0.09) \times 10^{-10}\,,
\end{equation}
where the next-to-leading order~(NLO) corrections of EW origin~\cite{Bobeth:2013tba}, as well as the QCD corrections up to the next-to-next-to-leading order~(NNLO)~\cite{Hermann:2013kca}, have been taken into account. Although the experimental uncertainties are still quite large, they are expected to get significantly reduced within the next few years~\cite{Bediaga:2012py}. All these experimental and theoretical progresses will lead to new stringent constraints on physics beyond the SM.

Motivated by the above considerations, in this work we shall perform a study of the rare leptonic decays $\Bqll$ within the A2HDM. Our paper is organized as follows. In section~\ref{sec:A2HDM} we give a brief overview of the A2HDM Lagrangian, especially of its Yukawa and scalar sectors. In section~\ref{sec:Theory} we summarize the SM results and describe the full one-loop calculation of the relevant Feynman diagrams in the A2HDM. We have performed the calculation in two different gauges, Feynman~($\xi=1$) and unitary~($\xi=\infty$), in order to check the gauge-independence of our results. In section~\ref{sec:numberical} we discuss the impact of the model parameters on the branching ratios of these decays, taking into account the latest implications from the LHC Higgs data. Our conclusions are made in section~\ref{sec:conclusion}. Finally, the appendix contains the explicit results for the individual Higgs-penguin diagrams.

\section{The aligned two-Higgs doublet model}
\label{sec:A2HDM}

The 2HDM extends the SM with the addition of a second scalar doublet of hypercharge $Y=\frac{1}{2}$~\cite{Lee:1973iz}. In the so-called ``Higgs basis", in which only one doublet gets a nonzero vacuum expectation value, the two doublets can be parametrized as
\begin{equation} \label{eq:Higgsbasis}
 \Phi_1=\left[ \begin{array}{c} G^+ \\ \frac{1}{\sqrt{2}}\, (v+S_1+iG^0) \end{array} \right] \; ,
 \qquad\qquad
 \Phi_2 = \left[ \begin{array}{c} H^+ \\ \frac{1}{\sqrt{2}}\, (S_2+iS_3)   \end{array}\right] \; ,
\end{equation}
where $G^\pm$ and $G^0$ denote the Goldstone fields, and $v=(\sqrt{2} G_F)^{-1/2} \simeq 246~\mathrm{GeV}$. The five physical scalar degrees of freedom are given by the two charged fields $H^\pm(x)$ and three neutral scalars $\varphi^0_i(x) =\{h(x), H(x), A(x)\}$. The latter are related with the $S_i$ fields through an orthogonal transformation, which is fixed by the scalar potential:
\beqn \label{eq:potential}
V &=& \mu_1\; \left(\Phi_1^\dagger\Phi_1\right)\, +\, \mu_2\; \left(\Phi_2^\dagger\Phi_2\right) \, +\, \left[\mu_3\; \left(\Phi_1^\dagger\Phi_2\right) \, +\, \mu_3^*\; \left(\Phi_2^\dagger\Phi_1\right)\right] \no\\[0.2cm]
& + & \lambda_1\, \left(\Phi_1^\dagger\Phi_1\right)^2 \, +\, \lambda_2\, \left(\Phi_2^\dagger\Phi_2\right)^2 \, +\, \lambda_3\, \left(\Phi_1^\dagger\Phi_1\right) \left(\Phi_2^\dagger\Phi_2\right) \, +\, \lambda_4\, \left(\Phi_1^\dagger\Phi_2\right) \left(\Phi_2^\dagger\Phi_1\right) \no\\[0.2cm]
& + & \left[\left(\lambda_5\; \Phi_1^\dagger\Phi_2 \, +\,\lambda_6\; \Phi_1^\dagger\Phi_1 \, +\,\lambda_7\; \Phi_2^\dagger\Phi_2\right) \left(\Phi_1^\dagger\Phi_2\right)\, +\, \mathrm{h.c.}\right]\,.
\eeqn

The Hermiticity of the potential requires all parameters to be real except $\mu_3$, $\lambda_5$, $\lambda_6$ and $\lambda_7$; thus, there are 14 real parameters. The minimization conditions $\langle 0|\Phi_1^T(x)|0\rangle = (0, v/\sqrt{2})$ and $\langle 0|\Phi_2^T(x)|0\rangle = (0, 0)$ impose the relations $\mu_1 = -\lambda_1 v^2$ and $\mu_3 = -\frac{1}{2}\,\lambda_6\, v^2$, which allow us to trade the parameters $\mu_1$ and $\mu_3$ by $v$ and $\lambda_6$, respectively. The freedom to rephase the field $\Phi_2$ implies, moreover, that only the relative phases among $\lambda_5$, $\lambda_6$ and $\lambda_7$ are physical. Therefore, we can fully characterize the potential with 11 parameters: $v$, $\mu_2$, $\lambda_{1,2,3,4}$, $|\lambda_{5,6,7}|$,
$\mathrm{arg}(\lambda_5\lambda_6^*)$ and  $\mathrm{arg}(\lambda_5\lambda_7^*)$. Four of these parameters can be determined through the physical scalar masses.

Inserting Eq.~(\ref{eq:Higgsbasis}) into Eq.~(\ref{eq:potential}), expanding out the resulting expression and imposing the minimization conditions, one can decompose the potential into a quadratic mass term plus cubic and quartic interactions~(up to an irrelevant constant). The mass term takes the form:
\beqn\label{eq:mass_term}
V_2 & = & M_{H^\pm}^2\, H^+ H^-\, +\, \frac{1}{2}\, \left(S_1, S_2, S_3\right)\; \mathcal{M}\; {\renewcommand{\arraystretch}{0.7}
\left(\ba S_1\\ S_2\\ S_3\ea\right)} \no \\[0.2cm]
& = & M_{H^\pm}^2\, H^+ H^-\, +\, \frac{1}{2}\, \sum_{i=1}^3\, M^2_{\varphi_i^0}\, \left(\varphi_i^0\right)^2\,,
\eeqn
with $M_{H^\pm}^2 = \mu_2 + \frac{1}{2}\lambda_3 v^2$ and
\begin{equation} \label{eq:mass_matrix}
\mathcal{M} \; = \; \left(\begin{array}{ccc}
2\lambda_1 v^2 & v^2\, \lambda_6^{\mathrm{R}} & -v^2\, \lambda_6^{\mathrm{I}}\\
v^2\, \lambda_6^{\mathrm{R}} & M_{H^\pm}^2  + v^2\left(\frac{\lambda_4}{2} + \lambda_5^{\mathrm{R}}\right)
& -v^2\, \lambda_5^{\mathrm{I}}\\
-v^2\, \lambda_6^{\mathrm{I}} & -v^2\, \lambda_5^{\mathrm{I}} & M_{H^\pm}^2  +  v^2\left(\frac{\lambda_4}{2} - \lambda_5^{\mathrm{R}}\right)
\ea\right)\, ,
\end{equation}
where $\lambda_i^{\mathrm{R}}\equiv \mathrm{Re}(\lambda_i)$ and $\lambda_i^{\mathrm{I}}\equiv \mathrm{Im}(\lambda_i)$. The symmetric mass matrix $\mathcal{M}$ is diagonalized by an orthogonal matrix $\mathcal{R}$, which defines the neutral mass eigenstates:
\begin{equation} \label{eq:mass_diagonalization}
\mathcal{R}\,\mathcal{M}\,\mathcal{R}^T\; =\; \mathrm{diag}\left( M_h^2, M_H^2, M_A^2\right)\,, \qquad\qquad
\varphi^0_i\; =\; \mathcal{R}_{ij}\, S_j\,.
\end{equation}
In a generic case, the three mass-eigenstates $\varphi^0_i(x)$ do not have definite CP quantum numbers.

In the CP-conserving limit, $\lambda_5^{\mathrm{I}}=\lambda_6^{\mathrm{I}} =\lambda_7^{\mathrm{I}} =0$ and $S_3$ does not mix with the other two neutral fields. The scalar spectrum contains then a CP-odd field $A=S_3$ and two CP-even scalars $h$ and $H$ which mix through the two-dimensional rotation matrix:\footnote{The scalar mixing is often parametrized in terms of $\alpha^{\prime}=\tilde\alpha +\pi/2$, so that the SM limit corresponds to $\alpha^{\prime}=\pi/2$~\cite{2HDM:review}. We prefer to describe small deviations from the SM limit with $\tilde\alpha\simeq0$.}
\begin{equation} \label{eq:CPC_mixing}
\left(\ba h\\ H\ea\right)\; = \;
\left[\bat \cos{\tilde\alpha} & \sin{\tilde\alpha} \\ -\sin{\tilde\alpha} & \cos{\tilde\alpha}\ea\right]\;
\left(\ba S_1\\ S_2\ea\right) \, .
\end{equation}
We shall adopt the conventions $M_h \le M_H$ and $0 \leq \tilde\alpha \leq \pi$, so that $\sin{\tilde\alpha}$ is always positive. The masses of the three physical neutral scalars are given in this case by
\begin{equation} \label{eq:CPC_masses}
M_h^2\; =\;\frac{1}{2}\,\left( \Sigma-\Delta\right)\, ,
\qquad
M_H^2\; =\;\frac{1}{2}\,\left( \Sigma+\Delta\right)\, ,
\qquad
M_A^2 \; =\; M_{H^\pm}^2\, +\, v^2\,\left(\frac{\lambda_4}{2} - \lambda_5^{\mathrm{R}}\right)\, ,
\end{equation}
where
\begin{equation} \label{eq:Sigma}
\Sigma \; =\; M_{H^\pm}^2\, +\, v^2\,\left(2\,\lambda_1 +\frac{\lambda_4}{2}+ \lambda_5^{\mathrm{R}}\right)\,,
\end{equation}
\begin{equation} \label{eq:Delta}
\Delta\; =\;\sqrt{\left[M_{H^\pm}^2\, +\, v^2\,\left(-2\,\lambda_1 +\frac{\lambda_4}{2}+ \lambda_5^{\mathrm{R}}\right) \right]^2 + 4 v^4 (\lambda_6^{\mathrm{R}})^2}
\; = \; - \frac{2 v^2\lambda_6^{\mathrm{R}}}{\sin{(2\tilde\alpha)}}\, ,
\end{equation}
and the mixing angle is determined through
\begin{equation} \label{eq:mixingCPC}
\tan{\tilde\alpha}\; =\; \frac{M_h^2 - 2\lambda_1 v^2}{v^2\lambda_6^{\mathrm{R}}}
\; =\; \frac{v^2\lambda_6^{\mathrm{R}}}{2\lambda_1 v^2- M_H^2}\, .
\end{equation}

The cubic and quartic self-couplings among the physical scalars and their interactions with the gauge bosons can be derived straightforwardly. Their explicit form could be found, for example, in Refs.~\cite{2HDM:review,A2HDM:collider2,2HDM:basisindep}.

\subsection{Yukawa sector}
\label{sec:Yukawa}

In the Higgs basis, the most generic Yukawa Lagrangian of the 2HDM is given by
\begin{equation} \label{eq:Yukawa1}
 \mathcal{L}_Y = -\frac{\sqrt{2}}{v}\,\Big[\bar{Q}'_L (M'_d \Phi_1 + Y'_d \Phi_2) d'_R + \bar{Q}'_L (M'_u \tilde{\Phi}_1 + Y'_u \tilde{\Phi}_2) u'_R + \bar{L}'_L (M'_\ell \Phi_1 + Y'_\ell \Phi_2) \ell'_R \Big] + \mathrm{h.c.} \,,
\end{equation}
where $\tilde{\Phi}_i(x)=i\tau_2\Phi_i^{\ast}(x)$ are the charge-conjugated scalar doublets with hypercharge $Y=-\frac{1}{2}$, $Q'_L$ and $L'_L$ denote the SM left-handed quark and lepton doublets, respectively, and $u'_R$, $d'_R$ and $\ell'_R$ are the corresponding right-handed singlets, in the weak interaction basis. All fermionic fields are written as $3$-vectors in flavour space and, accordingly, the couplings $M'_f$ and $Y'_f$~($f=u,d,\ell$) are $3\times3$ complex matrices.

In general, the Yukawa matrices $M_f'$ and $Y_f'$ cannot be simultaneously diagonalized in flavour space. Thus, in the fermion mass-eigenstate basis with diagonal mass matrices $M_f$, the corresponding Yukawa matrices $Y_f$ remain non-diagonal, giving rise to tree-level FCNC interactions. In the A2HDM, the tree-level FCNCs are eliminated by requiring the alignment in flavour space of the two Yukawa matrices coupling to a given type of right-handed fermions~\cite{Pich:2009sp}
\begin{equation} \label{eq:alignment}
 Y_{d,\ell}\; =\;\varsigma_{d,\ell}\, M_{d,\ell}\, ,
 \qquad\qquad
 Y_u\; =\;\varsigma^*_u\, M_u\, ,
\end{equation}
where the three proportionality parameters $\varsigma_f$~($f=d,u,\ell$) are arbitrary complex numbers and introduce new sources of CP violation. The Yukawa interactions of the physical scalars with the fermion mass-eigenstate fields then read~\cite{Pich:2009sp}
\begin{align} \label{eq:Yukawa2}
 \mathcal{L}_Y & = - \frac{\sqrt{2}}{v}\, H^+\, \Big\{\bar{u} \left[\varsigma_d\,V M_d P_R - \varsigma_u\,M_u^{\dagger} V P_L\right] d + \varsigma_\ell\,\bar{\nu} M_\ell P_R \ell \Big\} \nonumber \\[0.2cm]
 & \hspace{0.5cm} - \frac{1}{v}\; \sum_{\varphi^0_i,f}\, y^{\varphi^0_i}_f\, \varphi^0_i \, \left[\bar{f} M_f P_R f \right] + \mathrm{h.c.} \,,
\end{align}
where $P_{R,L}\equiv \frac{1\pm \gamma_5}{2}$ are the right-handed and left-handed chirality projectors, $M_f$ the diagonal fermion mass matrices, and $V$ the CKM quark-mixing matrix~\cite{CKM}. The couplings of the neutral scalar fields to fermion pairs are given by
\begin{equation} \label{eq:yukascal}
y_{d,\ell}^{\varphi^0_i} = \cR_{i1} + (\cR_{i2} + i\,\cR_{i3})\,\varsigma_{d,\ell} \,,
 \qquad\qquad
y_u^{\varphi^0_i} = \cR_{i1} + (\cR_{i2} - i\,\cR_{i3})\,\varsigma_{u}^* \,.
\end{equation}

In the A2HDM, all fermionic couplings to scalars are proportional to the corresponding fermion masses, and the only source of flavour-changing interactions is the CKM quark-mixing matrix $V$, while all leptonic couplings and the quark neutral-current interactions are diagonal in flavour. All possible freedom allowed by the alignment conditions is encoded by the three family-universal complex parameters $\varsigma_f$, which provide new sources of CP violation without tree-level FCNCs~\cite{Pich:2009sp}. The usual models with NFC, based on discrete $\mathcal{Z}_2$ symmetries, are recovered for particular values of the couplings $\varsigma_f$, as indicated in Table~\ref{tab:models}. Explicit examples of symmetry-protected underlying theories leading to a low-energy A2HDM structure have been discussed in Ref.~\cite{A2HDM:examples}.

\begin{table}[t]
  \centering
  \tabcolsep 0.30in
  \begin{tabular}{|c|c|c|c|}
  \hline
  Model   & $\varsigma_d$  & $\varsigma_u$ & $\varsigma_l$  \\
  \hline
  Type~I  & $ \cot{\beta}$ & $\cot{\beta}$ & $ \cot{\beta}$ \\
  Type~II & $-\tan{\beta}$ & $\cot{\beta}$ & $-\tan{\beta}$ \\
  Type~X (lepton-specific) & $ \cot{\beta}$ & $\cot{\beta}$ & $-\tan{\beta}$ \\
  Type~Y  (flipped) & $-\tan{\beta}$ & $\cot{\beta}$ & $ \cot{\beta}$ \\
  Inert   &      0         &        0      &       0        \\
  \hline
  \end{tabular}
  \caption{\label{tab:models} \small The one-to-one correspondence between different specific choices of the couplings $\varsigma_f$ and the 2HDMs based on discrete $\mathcal{Z}_2$ symmetries.}
\end{table}

The alignment conditions in Eq.~(\ref{eq:alignment}) presumably hold at some high-energy scale $\Lambda_A$ and are spoiled by radiative corrections. These higher-order contributions induce a misalignment of the Yukawa matrices, generating small FCNC effects suppressed by the corresponding loop factors~\cite{Pich:2009sp,Jung:2010ik,Pich:2010ic,A2HDM:RGE}. However, the flavour symmetries of the A2HDM tightly constrain the possible FCNC structures, keeping their effects well below the present experimental bounds~\cite{Jung:2010ik,A2HDM:flavour}. Using the renormalization-group equations~(RGEs)~\cite{A2HDM:RGE}, one can check that the only FCNC local structures induced at one loop take the form~\cite{Jung:2010ik,Pich:2010ic}
\begin{align} \label{eq:FCNCop}
 \mathcal{L}_{\mathrm{FCNC}} &= \frac{\cC}{4\pi^2 v^3}\, \left(1+\varsigma_u^*\,\varsigma_d\right)\, \sum_i\, \varphi^0_i \Big\{(\cR_{i2} + i\,\cR_{i3})\, (\varsigma_d-\varsigma_u)\, \left[\bar{d}_L\, V^{\dagger} M_u M_u^\dagger V M_d\, d_R\right] \nonumber \\
 & \hspace{4.2cm} - (\cR_{i2} - i\,\cR_{i3})\, (\varsigma_d^*-\varsigma_u^*)\, \left[\bar{u}_L\, V M_d M_d^\dagger V^\dagger M_u\, u_R\right] \Big\} + \mathrm{h.c.}\,,
\end{align}
which vanishes identically when $\varsigma_d = \varsigma_u$~($\mathcal{Z}_2$ models of types I, X and inert) or $\varsigma_d = -1/\varsigma_u^*$~(types II and Y), as it should be.

Although the numerical effect of the local term in Eq.~(\ref{eq:FCNCop}) is suppressed by $m_{q}m_{q'}^2/v^3$ and  quark-mixing factors, its tree-level contribution is needed to render finite the contribution from one-loop Higgs-penguin diagrams to $\Bqll$, as will be detailed later.
The renormalization of the coupling constant $\cC$ is determined to be
\begin{equation}\label{eq:Crenorm}
\cC = \cC_R(\mu) + \frac{1}{2}\,\left\{\frac{2\mu^{D-4}}{D-4} +\gamma_E-\ln{(4\pi)}\right\}\, ,
\end{equation}
where $D$ is the space-time dimension. Thus, the renormalized coupling satisfies
\begin{equation}
\cC_R(\mu) = \cC_R(\mu_0) - \ln{(\mu/\mu_0)}\,.
\end{equation}
Assuming the alignment to be exact at the scale $\Lambda_A$, \textit{i.e.}, $\cC_R(\Lambda_A) = 0$, this implies $\cC_R(\mu) = \ln{(\Lambda_A/\mu)}$.

\section{Calculation of $\boldsymbol{\overline{\mathcal{B}}(\Bqll)}$}
\label{sec:Theory}

\subsection{Effective Hamiltonian}
\label{sec:Heff}

The rare leptonic $\Bqll$ decays proceed through loop diagrams in both the SM and the A2HDM. After decoupling the heavy degrees of freedom, including the top quark, the weak gauge bosons, as well as the charged and neutral Higgs bosons, these decays are described by a low-energy effective Hamiltonian~\cite{Buras:2013uqa,DeBruyn:2012wk,Altmannshofer:2011gn}
\begin{equation} \label{eq:Heff}
 {\cal H}_{\rm eff}\; =\; -\frac{G_F\,\alpha}{\sqrt{2}\pi s^2_W}\left[
		V_{tb}^{\phantom{*}} V_{tq}^* \, \sum_{i}^{10,S,P} \left( C_i\,{\cal O}_i + C'_i\,{\cal O}'_i\right) + \mathrm{h.c.}\right]\,,
\end{equation}
where $G_F$ is the Fermi coupling constant, $\alpha=e^2/4\pi$ the QED fine-structure constant, and $s_W=\sin\theta_W$ the sine of the weak angle. The effective four-fermion operators are given, respectively, as
\begin{align} \label{eq:operators}
	{\cal O}_{10} &= (\bar q\gamma_\mu P_L b)\, (\bar \ell \gamma^\mu \gamma_5 \ell)\,, &
	{\cal O}'_{10} &= (\bar q\gamma_\mu P_R b)\, (\bar \ell \gamma^\mu \gamma_5 \ell)\,, \notag\\[0.2cm]	
	{\cal O}_{S} &= \frac{m_\ell m_b}{M^2_W}\; (\bar q P_R b)\, (\bar \ell\ell)\,, &
	{\cal O}'_{S} &= \frac{m_\ell m_b}{M^2_W}\; (\bar q P_L b)\, (\bar\ell\ell)\,, \notag\\[0.2cm]	
	{\cal O}_{P} &= \frac{m_\ell m_b}{M^2_W}\; (\bar q P_R b)\, (\bar\ell \gamma_5 \ell)\,, &
	{\cal O}'_{P} &= \frac{m_\ell m_b}{M^2_W}\; (\bar q P_L b)\, (\bar\ell \gamma_5 \ell)\,,
\end{align}
where $\ell=e,\mu,\tau$; $q=d,s$, and $m_b=m_b(\mu)$ denotes the $b$-quark running mass in the modified minimal subtraction~($\mathrm{\overline{MS}}$) scheme. In this paper, we shall neglect the operators ${\cal O}'_i$, because they only give contributions proportional to the light-quark mass $m_{q}$. Operators involving the vector current $\bar\ell \gamma^\mu\ell$ do not contribute to $\Bqll$ because the conserved vector current vanishes when contracted with the $B_q^0$ momentum. Since the matrix element $\langle0|\bar q\sigma_{\mu\nu} b|\bar B_q^0(p)\rangle=0$, there is also no contribution from the tensor operators. Thus, only the operators ${\cal O}_{10}$, ${\cal O}_{S}$ and ${\cal O}_{P}$ survive in our approximation.

As there are highly separated mass scales in the decays $\Bqll$, short-distance QCD corrections can contain large logarithms like $\ln{(\mu_b/M_W)}$ with $\mu_b\sim\mathcal{O}(m_b)$, which must be summed up to all orders in perturbation theory with the help of renormalization-group techniques. The evolution of the Wilson coefficients from the scale $\mathcal{O}(M_W)$ down to $\mathcal{O}(\mu_b)$ requires the solution of the RGEs of the corresponding operators ${\cal O}_{10}$, ${\cal O}_{S}$ and ${\cal O}_{P}$. However, the operator ${\cal O}_{10}$ has zero anomalous dimension due to the conservation of the $(V-A)$ quark current in the limit of vanishing quark masses. The operators ${\cal O}_{S}$ and ${\cal O}_{P}$ have also zero anomalous dimension, because the anomalous dimensions of the $b$-quark mass $m_b(\mu)$ and the scalar current $(\bar q P_R b)(\mu)$ cancel each other. Thus, with the operators defined by Eq.~(\ref{eq:operators}), the corresponding Wilson coefficients do not receive additional renormalization due to QCD corrections.

In the SM, the contributions from the scalar and pseudoscalar operators are quite suppressed and, therefore, are usually neglected in phenomenological analyses. However, they can be much more sizeable in models with enlarged Higgs sectors, such as the A2HDM, especially when the Yukawa and/or scalar-potential couplings are large. Therefore, the $\Bqll$ data provide useful constraints on the model parameters. To get the theoretical predictions for $\overline{\mathcal{B}}(\Bqll)$, the main task is then to calculate the three Wilson coefficients $C_{10,S,P}$ in both the SM and the A2HDM, details of which will be presented in the next few subsections.

\subsection{Computational method}
\label{sec:Method}

The standard way to find the Wilson coefficients is to require equality of one-particle irreducible amputated Green functions calculated in the full and in the effective theory~\cite{Buchalla:1995vs}. The former requires the calculation of various box, penguin and self-energy diagrams. We firstly use the program \texttt{FeynArts}~\cite{Hahn:2000kx}, with the model files provided by the package \texttt{FeynRules}~\cite{Christensen:2008py}, to generate all the Feynman diagrams contributing to the decays $\Bqll$, as well as the corresponding amplitudes, which can then be evaluated straightforwardly.

Throughout the whole calculation, we set the light-quark masses $m_{d,s}$ to zero; while for $m_b$, we keep it up to linear order. As the external momenta are much smaller than the masses of internal top-quark, gauge bosons, as well as charged and neutral scalars, the Feynman integrands are expanded in external momenta before performing the loop integration~\cite{Smirnov:1994tg}
\begin{equation} \label{eq:HME}
 \frac{1}{(k+l)^2-M^2} = \frac{1}{k^2-M^2}\,\left[1 - \frac{l^2+2(k\cdot l)}{k^2-M^2} + \frac{4(k\cdot l)^2}{(k^2-M^2)^2}\right]+\mathcal{O}(l^4/M^4)\,,
\end{equation}
where $k$ denotes the loop momentum, $M$ a heavy mass and $l$ an arbitrary external momentum. In addition, we employ the naive dimensional regularization scheme with an anti-commuting $\gamma_5$ to regularize the divergences appearing in Feynman integrals. After the Taylor expansion and factorizing out the external momenta, the integrals remain dependent only on the loop momentum and the heavy masses $M$. Subsequently, we apply the partial fraction decomposition~\cite{Bobeth:1999mk}
\begin{equation} \label{eq:PFD}
 \frac{1}{(q^2-m_1^2)(q^2-m_2^2)} = \frac{1}{m_1^2-m_2^2}\,\left[\frac{1}{q^2-m_1^2} - \frac{1}{q^2-m_2^2}\right]\,,
\end{equation}
which allows a reduction of all the Feynman integrals to those in which only a single mass parameter occurs in the propagator denominators. Finally, after reduction of tensor integrals to scalar ones, the only non-vanishing one-loop integrals take the form~\cite{Peskin:1995ev}
\begin{equation} \label{eq:scalarint}
 \int \frac{d^D k}{(2\pi)^D}\,\frac{1}{(k^2-m^2)^n} = \frac{(-1)^n i}{(4\pi)^{D/2}}\,\frac{\Gamma(n-D/2)}{\Gamma(n)}\, \left(\frac{1}{m^2}\right)^{n-D/2}\,,
\end{equation}
with an arbitrary integer power $n$ and with $m\neq0$.

The computational procedure has also been checked through an independent analytic calculation of the Feynman diagrams, using more standard techniques such as the Feynman parametrization to combine propagators. We found full agreement between the results obtained with these two methods.

It should be noted that, in deriving the effective Hamiltonian in Eq.~(\ref{eq:Heff}), the limit $m_{u,c}\to 0$ and the unitarity of the CKM matrix,
\begin{equation} \label{eq:ckmunitary}
V_{uq}^* V_{ub}^{\phantom{*}} + V_{cq}^* V_{cb}^{\phantom{*}} + V_{tq}^* V_{tb}^{\phantom{*}}\; =\; 0\, ,
\end{equation}
have been implicitly exploited. In general, the Wilson coefficients $C_i$ are functions of the internal up-type quark masses, together with the corresponding CKM factors~\cite{Buchalla:1995vs}:
\begin{equation}
C_i\; =\; \sum_{j=u,c,t}V_{jq}^* V_{jb}^{\phantom{*}}\; F_i(x_j)\, ,
\end{equation}
where $x_j=m_j^2/M_W^2$, and $F_i(x_j)$ denote the loop functions. The unitarity relation in Eq.~(\ref{eq:ckmunitary}) implies vanishing coefficients $C_i$ if the internal quark masses are set to be equal, \textit{i.e.}, $x_u=x_c=x_t$. For this reason, we need only to calculate explicitly the contributions from internal top quarks, while those from up and charm quarks are taken into account by means of simply omitting the mass-independent terms in the basic functions $F_i(x_t)$. For simplicity, we also introduce the following mass ratios:
\begin{eqnarray}
x_t = \frac{m^2_t}{M^2_W}, \qquad  x_{H^+} = \frac{M^2_{H^\pm}}{M^2_W}, \qquad
x_{\varphi^0_i} =  \frac{M^2_{\varphi^0_i}}{M^2_W},
\qquad
x_{h_{\rm SM}} =  \frac{M^2_{h_{\rm SM}}}{M^2_W}\,,
\end{eqnarray}
where $m_t=m_t(\mu)$ is the top-quark running mass in the $\mathrm{\overline{MS}}$ scheme, and $h_{\rm SM}$ the SM Higgs boson.

In order to make a detailed presentation of our results, we shall split the different contributions to the Wilson coefficients into the form:
\begin{eqnarray}
C_{10} & = & C^{\rm SM}_{10} \, +\,  C^{\rm Z\, penguin,\, \rm A2HDM}_{10} \, ,
\\[0.2cm]
C_{S} &=&  C^{\rm box,\, \rm SM}_{S}\, +\,  C^{\rm box,\, \rm A2HDM}_{S}\, +\,
C^{\varphi_i^0,\, \rm A2HDM}_{S}\, ,
\\[0.2cm]
C_{P} &=&  C^{\rm box,\, \rm SM}_{P}\, +\, C^{\rm Z\, penguin,\, \rm SM}_{P}\, +\, C^{\rm GB\, penguin,\, \rm SM}_{P} \, +\,  C^{\rm box,\, \rm A2HDM}_{P}
\nonumber \\[0.2cm]
&&  +\; C^{\rm Z\, penguin,\, \rm A2HDM}_{P}\, +\, C^{\rm GB\, penguin,\, \rm A2HDM}_{P}
\, +\, C^{\varphi_i^0,\, \rm A2HDM}_{P}\, .
\end{eqnarray}
The pieces labeled with ``SM" only involve SM fields~(without the Higgs), while those denoted by ``A2HDM" contain the scalar contributions. We have calculated all the individual diagrams in both the Feynman~($\xi=1$) and the unitary~($\xi=\infty$) gauges. Goldstone boson~(GB) contributions are of course absent in the unitary gauge. While the contributions of the box and penguin diagrams to the Wilson coefficients are separately gauge dependent, their sum is indeed independent of the EW gauge fixing~\cite{Buchalla:1990qz,Botella:1986gf}. Note that photonic penguin diagrams, in both the SM and the A2HDM, do not contribute to the decays $\Bqll$ because of the pure vector nature of the electromagnetic leptonic coupling.

In $\Bqll$ the external momenta are small compared to the EW scale $M_W$. One can then set all external momenta to zero when evaluating $C_{10}$. However, the external momenta must be taken into account to evaluate the scalar Wilson coefficients $C_S$ and $C_P$, otherwise some contributions would be missed.

\subsection{Wilson coefficients in the SM}
\label{sec:WCSM}

\begin{figure}[t]
  \centering
  \includegraphics[width=15cm]{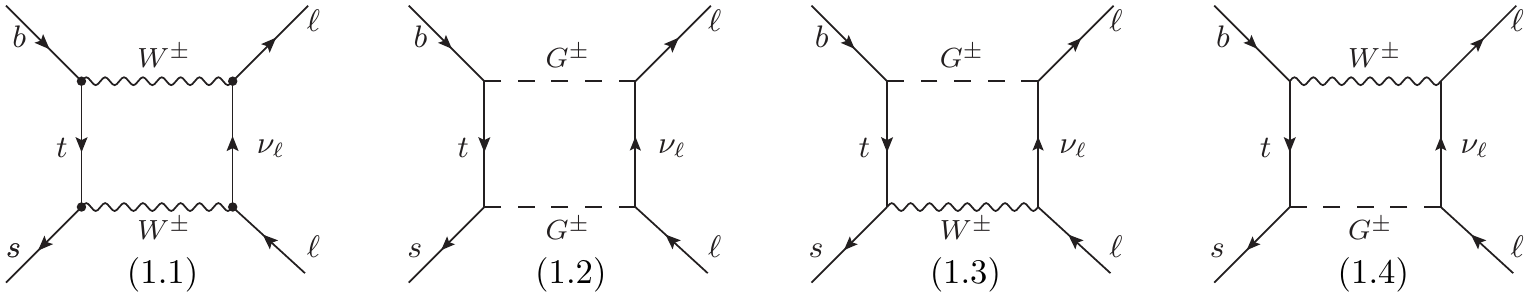}
  \caption{\small SM $W$-box diagrams contributing to $\bar{B}^0_s\to \ell^+\ell^-$.
  Diagrams involving Goldstone bosons $G^{\pm}$ are absent in the unitary gauge.}
  \label{fig:BoxSM}
\end{figure}

\begin{figure}[t]
  \centering
  \includegraphics[width=15cm]{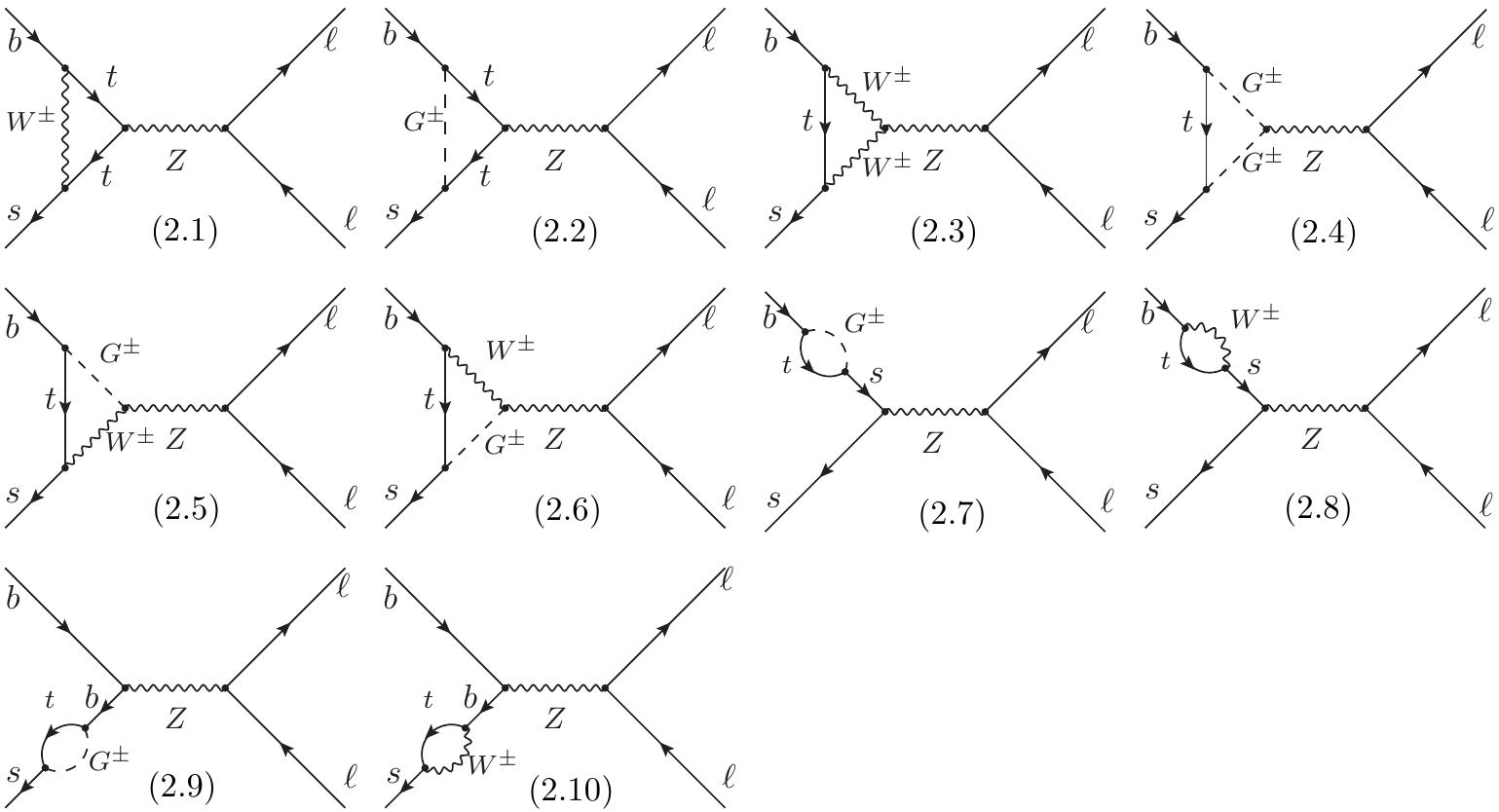}
  \caption{\small SM $Z$-penguin diagrams contributing to $\bar{B}^0_s\to \ell^+\ell^-$.
  Diagrams involving Goldstone bosons $G^{\pm}$ are absent in the unitary gauge.}
  \label{fig:ZSM}
\end{figure}

In the SM, the dominant contributions to the decays $\Bqll$ come from the $W$-box and $Z$-penguin diagrams shown in Figs.~\ref{fig:BoxSM} and \ref{fig:ZSM}, respectively, which generate the Wilson coefficient:
\begin{equation}
 C^{\rm{SM}}_{10} \;=\; - \eta_Y^{\rm EW}\,\eta_Y^{\rm QCD}\,Y_0(x_t)\,,
 \label{eq:C10SM}
\end{equation}
where
\begin{equation}\label{eq:Y0}
Y_0(x_t) \, =\, \frac{x_t}{8}\left[ \frac{x_t-4}{x_t-1} + \frac{3x_t}{(x_t-1)^2}\ln x_t \right]
\end{equation}
is the one-loop function that was calculated for the first time in Ref.~\cite{Inami:1980fz}. The factor $\eta_Y^{\rm EW}$ accounts for both the NLO EW matching corrections~\cite{Bobeth:2013tba}, as well as the logarithmically enhanced QED corrections that originate from the renormalization group evolution~\cite{Bobeth:2013uxa,Hermann:2013kca}, while the coefficient $\eta_Y^{\rm QCD}$ stands for the NLO~\cite{Buchalla:1998ba,Misiak:1999yg} and NNLO~\cite{Hermann:2013kca} QCD corrections.

When the small external momenta are taken into account, the SM $W$-box and $Z$-penguin diagrams also generate contributions to the Wilson coefficients $C_{S}$ and $C_P$. The contribution from diagram \ref{fig:BoxSM}.2 can be neglected, because it contains two leptonic Goldstone couplings which generate a suppression factor $m^2_\ell/M^2_W$. The scalar contribution from the remaining box diagrams is given by:
\begin{eqnarray}
 C^{\rm box,\, \rm SM}_{S, \,\rm Feynman} &=&- \frac{x_t(x_t-2)}{12(x_t-1)^2} + \frac{(x_t-2)(3x_t-1)}{24(x_t-1)^3} \,\ln x_t \,,\\[0.2cm]
 C^{\rm box,\, \rm SM}_{S, \rm Unitary} &=& -\frac{x_t(x_t+1)}{48(x_t-1)^2} - \frac{(x_t-2)(3x_t^2-3x_t+1)}{24(x_t-1)^3}\, \ln x_t\, ,
 \label{eq:csSM}
\end{eqnarray}
where the two different expressions correspond to the results obtained in the Feynman and unitary gauges, respectively.

\begin{figure}[t]
  \centering
  \includegraphics[width=15cm]{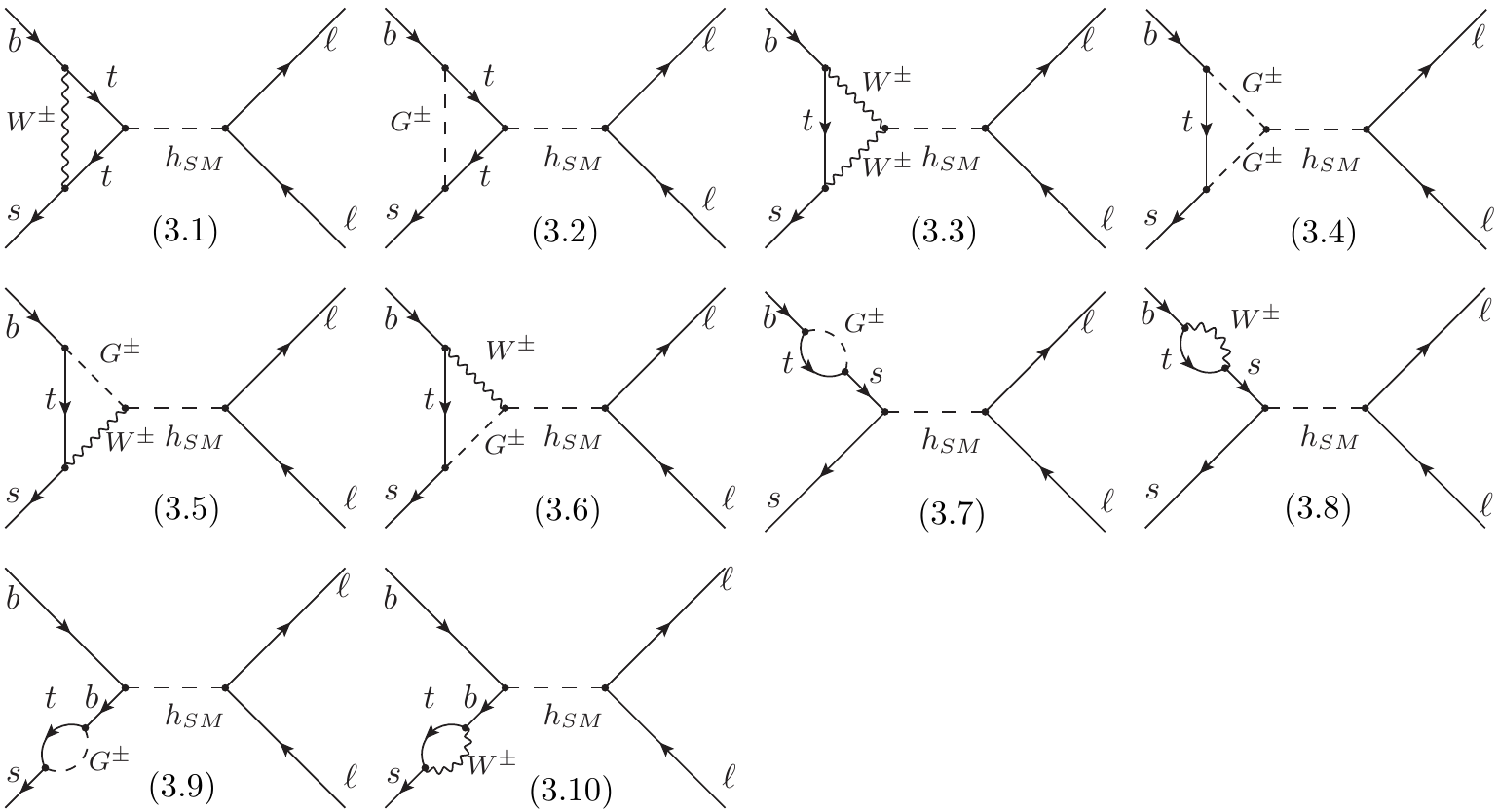}
  \caption{\small SM Higgs-penguin diagrams contributing to $\bar{B}^0_s\to \ell^+\ell^-$. Contributions with Goldstone bosons $G^{\pm}$ are absent in the unitary gauge.}
  \label{fig:hSM}
\end{figure}

In the SM there is an additional contribution to the scalar Wilson coefficient $C_S$ from the Higgs-penguin diagrams shown in Fig.~\ref{fig:hSM}, which is by itself gauge dependent~\cite{Botella:1986gf,Grzadkowski:1983yp,Krawczyk:1989qp} and should cancel the gauge dependence of the $W$-box contribution. We find the result:
\begin{eqnarray} \label{eq:csSM1}
 C^{\rm h\, penguin,\, \rm SM}_{S, \,\rm Feynman} &=&- \frac{x_t}{8}\, \left[\frac{3}{x_{h_{\rm SM}}}-\frac{x_t-3}{2 (x_t-1)^2} + \frac{x_t(x_t-2)}{(x_t-1)^3}\, \ln x_t \right]\,,\\[0.2cm]
 C^{\rm h\, penguin,\, \rm SM}_{S, \,\rm Unitary} &=&- \frac{3 x_t}{8 x_{h_{\rm SM}}}\,.
   \label{eq:csSM2}
\end{eqnarray}
The sum of the two contributions to $C_S$ is indeed gauge independent:
\begin{align} \label{eq:csSMfinal}
C^{\rm SM}_{S} &\; =\; C^{\rm box,\, \rm SM}_{S, \,\rm Feynman}
+C^{\rm h\, penguin,\, \rm SM}_{S, \rm Feynman}\; =\;
C^{\rm box,\, \rm SM}_{S, \,\rm Unitary}+C^{\rm h\, penguin,\, \rm SM}_{S, \,\rm Unitary}
\nonumber\\[0.2cm]
&\; =\; -\frac{3x_t}{8 x_{h_{\rm SM}}} - \frac{x_t(x_t+1)}{48(x_t-1)^2} - \frac{(x_t-2)(3x_t^2-3x_t+1)}{24(x_t-1)^3}\, \ln x_t \, .
\end{align}

The contribution from the SM $W$-box diagrams~(Fig.~\ref{fig:BoxSM}) to the pseudoscalar
Wilson coefficient $C_P$ is given by:
\begin{eqnarray}\label{eq:cpSM11}
 C^{\rm box,\, \rm SM}_{P, \,\rm Feynman} &=& \frac{x_t(35x_t^2-82x_t-1)}{72(x_t-1)^3} - \frac{9x_t^3-28x_t^2+x_t+2}{24(x_t-1)^4}\, \ln x_t \,,\\[0.2cm]
 C^{\rm box,\, \rm SM}_{P, \,\rm Unitary} &=& \frac{x_t(71x_t^2-172 x_t-19)}{144(x_t-1)^3} + \frac{x_t^4-12x_t^3+34 x_t^2-x_t-2}{24(x_t-1)^4} \,\ln x_t \,.
 \label{eq:cpSM12}
\end{eqnarray}

\begin{figure}[t]
  \centering
  \includegraphics[width=15cm]{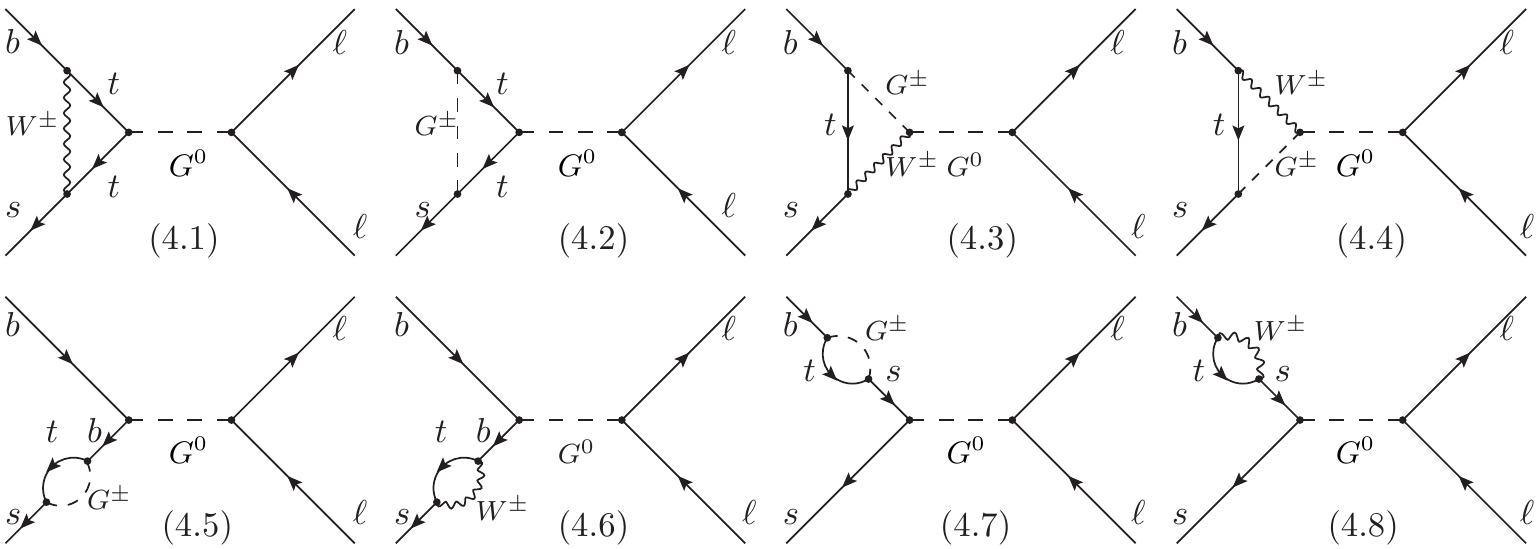}
  \caption{\small SM Goldstone-penguin diagrams contributing to $\bar{B}^0_s\to \ell^+\ell^-$. These contributions are absent in the unitary gauge.}
  \label{fig:GBSM}
\end{figure}

Additional contributions to $C_P$ are generated by the $Z$- and Goldstone-penguin diagrams shown in Figs.~\ref{fig:ZSM} and \ref{fig:GBSM}, respectively. The contributions from diagrams \ref{fig:GBSM}.6, \ref{fig:GBSM}.7 and \ref{fig:GBSM}.8 are proportional to the light-quark mass and can be therefore neglected. We find:
\begin{eqnarray} \label{eq:cpSM2}
 C^{\rm Z\, penguin,\, \rm SM}_{P, \,\rm Feynman} &=& \frac{x_t(5 x_t^2+16 x_t+3)}{48 (x_t-1)^3} - \frac{x_t^4+x_t^3+18 x_t^2-12 x_t+4}{24(x_t-1)^4}\, \ln x_t \nonumber \\[0.2cm]
 &&\mbox{} - s_W^2\,\left[\frac{x_t(5 x_t^2+40 x_t-21)}{72 (x_t-1)^3} - \frac{3 x_t^4-3 x_t^3+36 x_t^2-32 x_t+8}{36(x_t-1)^4} \ln x_t\right]\,,\\[0.3cm]
 C^{\rm GB\, penguin,\, \rm SM}_{P, \,\rm Feynman} &=& \left(1-s_W^2\right)\,\frac{x_t}{4}\,\left[\frac{x_t-6}{x_t-1} + \frac{3 x_t+2}{(x_t-1)^2} \,\ln x_t \right]\, ,
 \label{eq:cpZGBSM}
 \end{eqnarray}
and
\begin{eqnarray} \label{eq:cpSM3}
 C^{\rm Z\, penguin,\, \rm SM}_{P, \,\rm Unitary} &=& \frac{1}{12}\,\left[\frac{x_t(18 x_t^3-137 x_t^2+262 x_t-95)}{6 (x_t-1)^3} + \frac{8 x_t^4-11 x_t^3-15 x_t^2+12 x_t-2}{(x_t-1)^4}\, \ln x_t \right]
\nonumber \\[0.2cm]
&& \hspace{-1.75cm}\mbox{}
- \frac{s_W^2}{36}\,\left[\frac{x_t(18 x_t^3-139 x_t^2+274 x_t-129)}{2 (x_t-1)^3} + \frac{24 x_t^4-33 x_t^3-45 x_t^2+50 x_t-8}{(x_t-1)^4}\, \ln x_t\right].
\end{eqnarray}
Using the above results, one can easily check that the SM contribution to $C_P$ is also gauge independent:
\begin{align} \label{eq:cpSM}
C^{\rm SM}_{P} &\; =\; C^{\rm box,\, \rm SM}_{P, \,\rm Feynman} +C^{\rm Z\, penguin,\, \rm SM}_{P, \rm Feynman} +C^{\rm GB\, penguin,\, \rm SM}_{P, \,\rm Feynman}
\; =\; C^{\rm box,\, \rm SM}_{P, \rm Unitary} +C^{\rm Z\, penguin,\, \rm SM}_{P, \rm
\, Unitary}
\nonumber \\[0.2cm]
&\; =\; \frac{1}{24}\,\left[\frac{x_t(36 x_t^3-203 x_t^2+352 x_t-209)}{6 (x_t-1)^3} + \frac{17 x_t^4-34 x_t^3+ 4 x_t^2+23 x_t-6}{(x_t-1)^4} \,\ln x_t \right] \nonumber
\\[0.2cm]
&\; -\; \frac{s_W^2}{36}\,\left[\frac{x_t(18 x_t^3-139 x_t^2+274 x_t-129)}{2 (x_t-1)^3} + \frac{24 x_t^4-33 x_t^3-45 x_t^2+50 x_t-8}{(x_t-1)^4}\, \ln x_t\right].
\end{align}

The GIM mechanism has eliminated those contributions which are independent of the virtual top-quark mass. However, the $\ln{x_t}$ terms in the Wilson coefficients $C_S^{\rm SM}$ and $C_P^{\rm SM}$ do not vanish in the massless limit: at $x_t\ll 1$, $C_S^{\rm SM}\sim -\frac{1}{12}\,\ln{x_t}$ and $C_P^{\rm SM}\sim -\frac{1}{4}\,\left( 1-\frac{8}{9}\, s_W^2\right)\,\ln{x_t}$. These infrared-sensitive terms arise from diagrams 1.1 and 2.1 in both gauges. The corresponding contributions from virtual up and charm quarks cancel in the matching process with the low-energy effective theory, which has the same infrared behaviour.\footnote{
In the low-energy effective theory the same $\ln{x_c}$ ($\ln{x_u}$) terms appear from
analogous diagrams with a $c\,\bar\nu_\ell$ ($u\,\bar\nu_\ell$) or $c\,\bar c$ ($u\,\bar u$) loop connecting two four-fermion operators.}

\subsection{Wilson coefficients in the A2HDM}
\label{sec:WCA2HDM}

\begin{figure}[t]
  \centering
  \includegraphics[width=15cm]{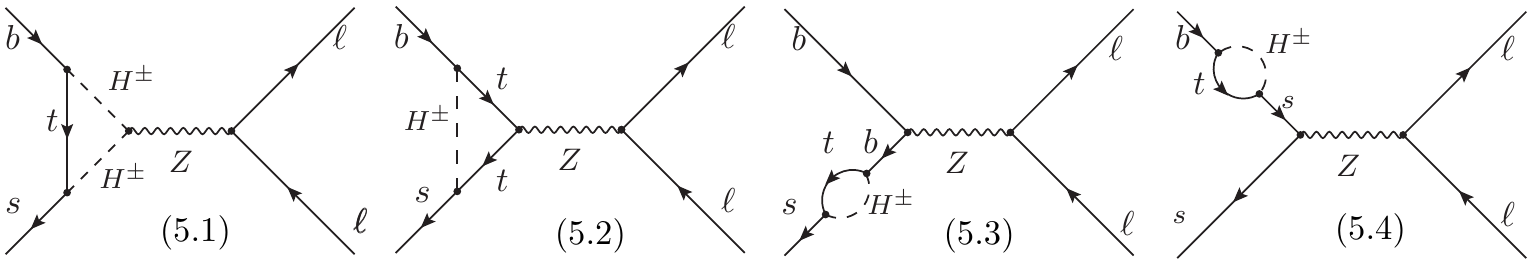}
  \caption{\small $Z$-penguin diagrams involving $H^\pm$ exchanges in the A2HDM.}
  \label{fig:ZA2HDM}
\end{figure}

In the A2HDM, the only new contribution to $C_{10}$ comes from the $Z$-penguin diagrams shown in Fig.~\ref{fig:ZA2HDM}. The result is gauge independent and given by
\begin{equation} \label{eq:C10A2HDM}
C^{\rm A2HDM}_{10}\; =\; C^{\rm Z\, penguin,\, \rm A2HDM}_{10}\; =\; |\varsigma_u|^2\,\frac{x_t^2}{8}\, \left[\frac{1}{x_{H^+}-x_t} + \frac{x_{H^+}}{(x_{H^+}-x_t)^2}\,\left(\ln x_t - \ln x_{H^+}\right)\right]\,.
\end{equation}
In the particular case of the type-II 2HDM (or MSSM), $\varsigma_u=1/\tan\beta$, this result agrees with the one calculated in Ref.~\cite{Chankowski:2000ng}.

\begin{figure}[t]
  \centering
  \includegraphics[width=15cm]{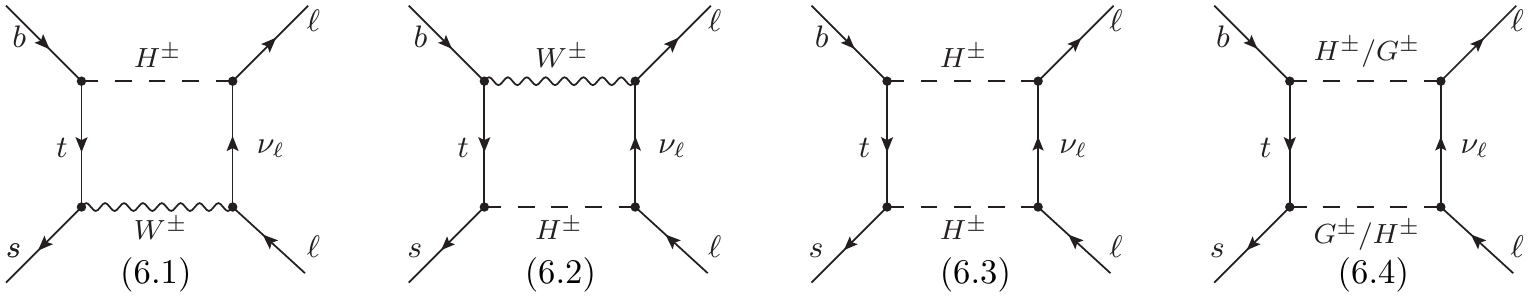}
  \caption{\small Box diagrams involving $H^\pm$ exchanges in the A2HDM. Diagrams with Goldstone bosons are absent in the unitary gauge.}
  \label{fig:BoxA2HDM}
\end{figure}

The box diagrams shown in Fig.~\ref{fig:BoxA2HDM} involve charged scalar exchanges and contribute to the Wilson coefficients $C^{\rm A2HDM}_S$ and $C^{\rm A2HDM}_P$. The contributions from diagrams \ref{fig:BoxA2HDM}.3 and \ref{fig:BoxA2HDM}.4 can be neglected, since they are proportional to $m^2_\ell/M^2_W$.  For the scalar coefficients we find the results:
\begin{align}
 C^{\rm box,\, \rm A2HDM}_{S, \,\rm Feynman} &\; =\;
 \frac{x_t}{8 (x_{H^+}-x_t)}\,\Bigg\{
 \varsigma_{\ell}\,\varsigma_u^*\, \left[\frac{x_t-x_{H^+}}{(x_{H^+}-1)(x_t-1)} + \frac{x_t}{(x_t-1)^2}\,\ln x_t - \frac{x_{H^+}}{(x_{H^+}-1)^2}\,\ln x_{H^+}\right] \nonumber\\[0.2cm]
 &\;\mbox{} -\varsigma_u\,\varsigma^*_{\ell}\left[\frac{1}{x_{H^+}-1} + \frac{ x_{H^+}}{(x_{H^+}-x_t)(x_t-1)}\,\ln x_t - \frac{x_{H^+}(2x_{H^+}-x_t-1)}{(x_{H^+}-x_t)(x_{H^+}-1)^2}\,\ln x_{H^+}\right] \nonumber\\[0.2cm]
 &\;\mbox{} + 2\,\varsigma_d\,\varsigma_{\ell}^*\,\left[\frac{1}{x_{H^+}-1}\,\ln x_{H^+} - \frac{1}{x_t-1}\,\ln x_t\right]\Bigg\}\,,
\label{eq:CSFeynmanboxA2HDM} \\[0.3cm]
 C^{\rm box,\, \rm A2HDM}_{S, \,\rm Unitary} &\; =\;
 \frac{x_t}{8 (x_{H^+}-x_t)}\,\Bigg\{\varsigma_{\ell}\,\varsigma_u^*\, \left[ \frac{x_t}{x_t-1}\,\ln x_t - \frac{x_{H^+}}{x_{H^+}-1}\,\ln x_{H^+}\right]
 \nonumber\\[0.2cm]
 &\;\mbox{} + \varsigma_u\,\varsigma_{\ell}^*\,\left[1- \frac{x_{H^+}-x_t^2}{(x_{H^+}-x_t)(x_t-1)}\,\ln x_t - \frac{x_{H^+}(x_t-1)}{(x_{H^+}-x_t)(x_{H^+}-1)}\,\ln x_{H^+}\right]
 \nonumber\\[0.2cm]
 &\;\mbox{} + 2\,\varsigma_d\,\varsigma_{\ell}^*\,
 \Big[\ln x_t - \ln x_{H^+}\Big]\Bigg\}\, ,
\label{eq:CSUnitaryboxA2HDM}
\end{align}
while the pseudoscalar contributions are given by:
\begin{align}
 C^{\rm box,\, \rm A2HDM}_{P, \,\rm Feynman} &\;=\; \left. - C^{\rm box,\, \rm A2HDM}_{S, \,\rm Feynman}\right|_{\varsigma_{\ell}\, \varsigma_u^* \to -\varsigma_{\ell}\, \varsigma_u^*}\,,\\[0.2cm]
 C^{\rm box,\, \rm A2HDM}_{P, \,\rm Unitary} &\;=\; \left. - C^{\rm box,\, \rm A2HDM}_{S, \,\rm Unitary}\right|_{\varsigma_{\ell}\, \varsigma_u^* \to -\varsigma_{\ell}\, \varsigma_u^*}\, .
 \label{eq:CPboxA2HDM}
\end{align}
Most of the previous calculations in the literature focused on the type-II 2HDM in the large $\tan\beta$ limit; \textit{i.e.}, only those contributions proportional to $\tan^2{\beta}$ were kept, which correspond to the $\varsigma_d\,\varsigma^*_\ell$ terms in Eqs.~(\ref{eq:CSFeynmanboxA2HDM})--(\ref{eq:CPboxA2HDM}). For this specific case, our results agree with Ref.~\cite{Logan:2000iv}.

\begin{figure}[t]
  \centering
  \includegraphics[width=13cm]{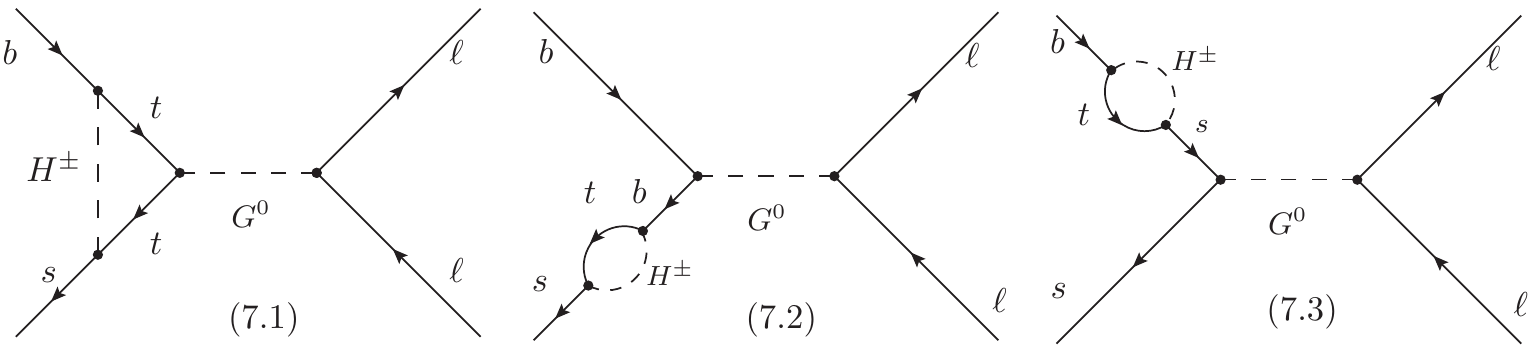}
  \caption{\small Goldstone-boson penguin diagrams involving $H^\pm$ exchanges in the A2HDM. These contributions are absent in the unitary gauge.}
  \label{fig:GBA2HDM}
\end{figure}

Similarly to the SM case, the coefficient $C^{\rm A2HDM}_P$ receives additional contributions from $Z$- and Goldstone-penguin diagrams shown in Figs.~\ref{fig:ZA2HDM} and \ref{fig:GBA2HDM}, respectively. They are given by:
\begin{align}
 C^{\rm Z\, penguin,\, \rm A2HDM}_{P, \,\rm Feynman} &\; =\;
 \frac{x_t}{4(x_{H^+}-x_t)^2}\, \Bigg\{\varsigma_{d}\, \varsigma_u^*\,\left[- \frac{x_t+x_{H^+}}{2} + \frac{x_t x_{H^+}}{x_{H^+}-x_t}\, (\ln x_{H^+}-\ln x_t)\right] \nonumber\\[0.2cm]
 & \hspace{-0.8cm}\mbox{} + |\varsigma_u|^2\,\frac{1}{6(x_{H^+}-x_t)}\,\left[\frac{x_{H^+}^2-8x_{H^+}x_t-17x_t^2}{6} + \frac{x_t^2(3x_{H^+}+x_t)}{x_{H^+}-x_t}\, (\ln x_{H^+}-\ln x_t)\right]\Bigg\} \nonumber\\[0.2cm]
 & \hspace{-1.8cm}\mbox{} + s_W^2\,\frac{x_t}{6(x_{H^+}-x_t)^2}\,\Bigg\{\varsigma_{d}\, \varsigma_u^*\,\left[\frac{5 x_t-3 x_{H^+}}{2} + \frac{x_{H^+}(2x_{H^+}-3x_t)}{x_{H^+}-x_t}\, (\ln x_{H^+}-\ln x_t)\right] \nonumber\\[0.2cm]
 & \hspace{-0.7cm}\mbox{} - |\varsigma_u|^2\,\frac{1}{6(x_{H^+}-x_t)}\,\left[\frac{4 x_{H^+}^3-12 x_{H^+}^2 x_t+9 x_{H^+} x_t^2+3 x_t^3}{x_{H^+}-x_t}\, (\ln x_{H^+}-\ln x_t) \right. \nonumber\\[0.2cm]
 & \left. \hspace{4cm}\mbox{} - \frac{17 x_{H^+}^2-64 x_{H^+}x_t+71 x_t^2}{6} \right]\Bigg\}\,,
\label{eq:CPFeynmanZA2HDM} \\[0.2cm]
 C^{\rm GB\, penguin,\, \rm A2HDM}_{P, \,\rm Feynman} &\; =\; |\varsigma_u|^2\,(1-s_W^2)\,\frac{x_t^2}{4 (x_{H^+}-x_t)^2}\,\Big[x_{H^+}\, (\ln x_{H^+}-\ln x_t)+x_t-x_{H^+}\Big]\,.
\label{eq:CPFeynmanGBA2HDM}
\end{align}
The gauge dependence of these two contributions compensates each other. Since there is no contribution from Goldstone-penguin topologies in the unitary gauge, the $Z$-penguin result should satisfy in this case:
\begin{equation}
C^{\rm Z\, penguin,\, \rm A2HDM}_{P, \,\rm Unitay} =
C^{\rm Z\, penguin,\, \rm A2HDM}_{P, \,\rm Feynman}+C^{\rm GB\, penguin,\, \rm A2HDM}_{P, \,\rm Feynman}\, .
\label{eq:CPUnitaryZA2HDM}
\end{equation}
This relation has been validated by the actual calculation.

\subsubsection{Neutral scalar exchange}
\label{sec:hiA2HDM}

\begin{figure}[t]
  \centering
  \includegraphics[width=4.2cm]{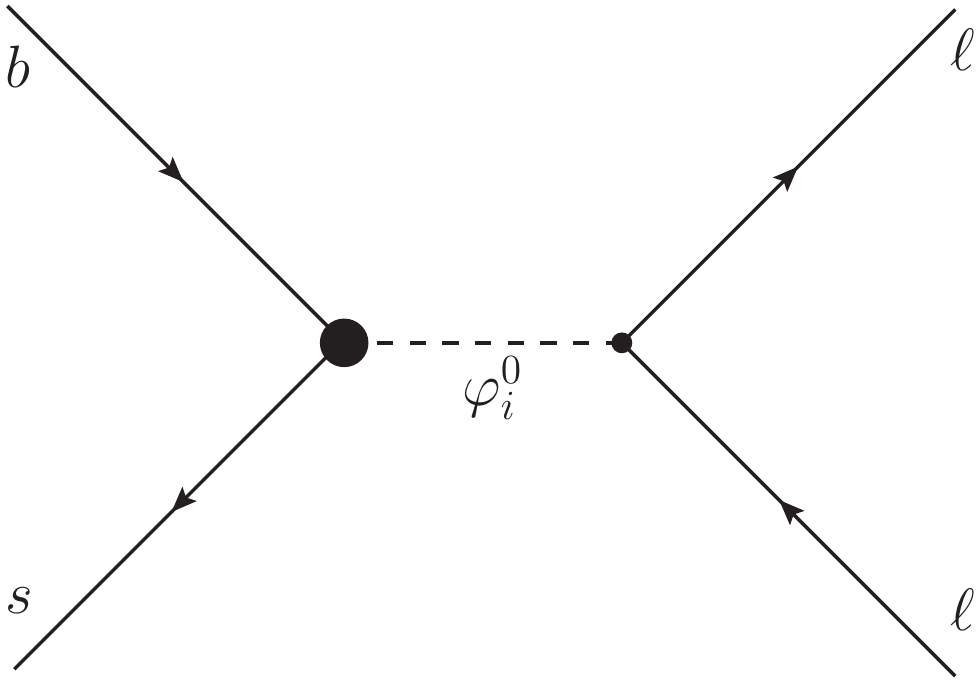}
  \caption{\small Tree-level FCNC diagram mediated by the neutral scalars $\varphi^0_i=\{h,H,A\}$.}
  \label{fig:hitree}
\end{figure}

\begin{figure}[t]
  \centering
  \includegraphics[width=15cm]{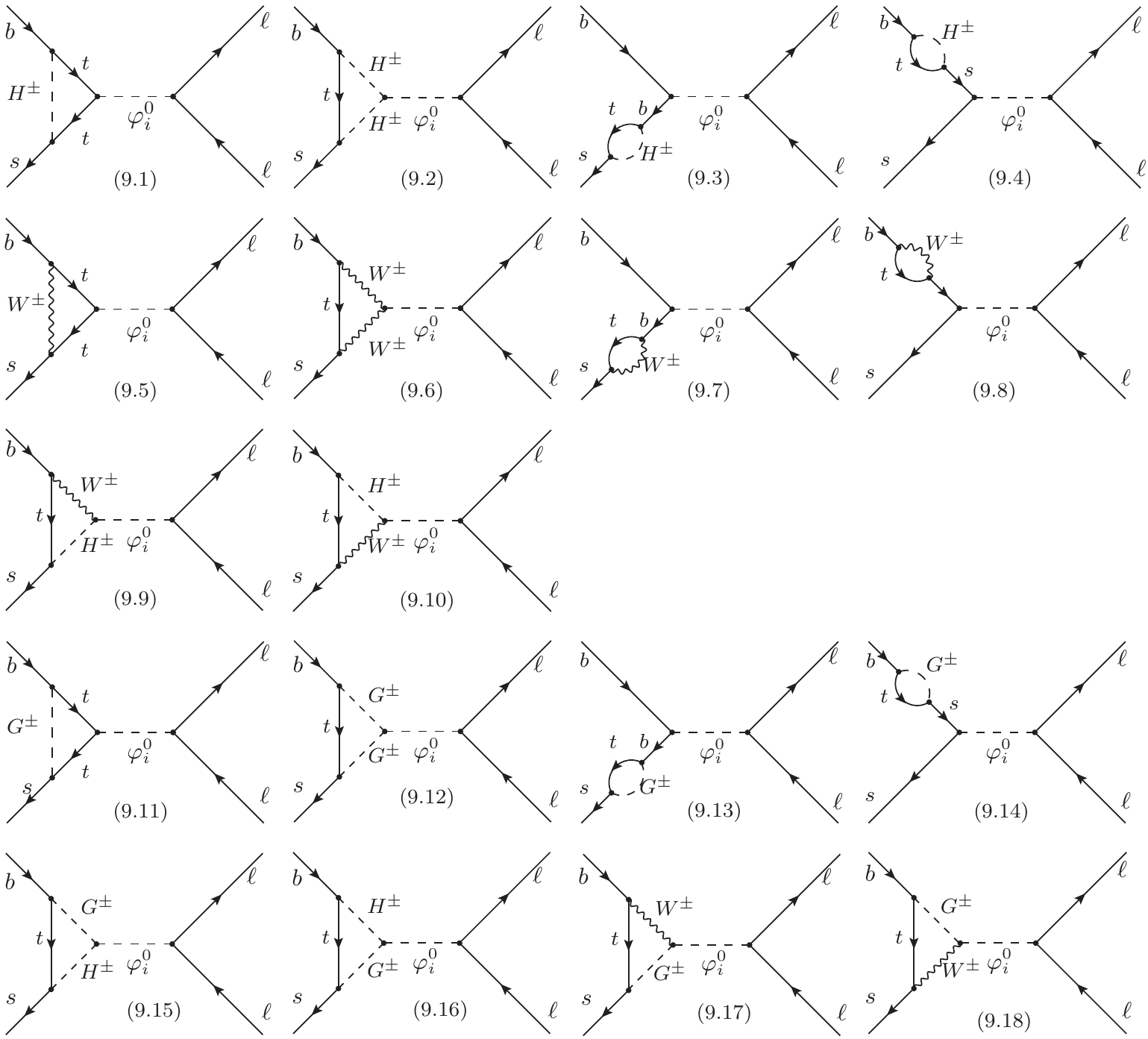}
  \caption{\small Scalar penguin diagrams in the A2HDM, where $\varphi^0_i=\{h,H,A\}$. Diagrams \ref{fig:hiA2HDM}.11 to \ref{fig:hiA2HDM}.18 are absent in the unitary gauge.}
  \label{fig:hiA2HDM}
\end{figure}

The Wilson coefficients $C^{\rm A2HDM}_S$ and $C^{\rm A2HDM}_P$ receive a direct tree-level contribution from the scalar-exchange diagram shown in Fig.~\ref{fig:hitree}, where the FCNC vertex $\varphi_i^0 \bar s b$ is generated by the local operator in Eq.~\eqn{eq:FCNCop}. This contribution must be combined together with the scalar penguin diagrams shown in Fig.~\ref{fig:hiA2HDM}. The structure of the common $\varphi_i^0 \bar\ell\ell$ vertex relates the resulting scalar and pseudoscalar Wilson coefficients, which take the form:
\begin{equation} \label{eq:ScalarPenguinWCs}
C_S^{\varphi_i^0,\, \rm A2HDM}\; = \; \sum_{\varphi_i^0}\; \mathrm{Re} (y_\ell^{\varphi^0_i})
\; \hat{C}^{\varphi_i^0}\, ,
\qquad\qquad
C_P^{\varphi_i^0,\, \rm A2HDM}\; = \; i\;\sum_{\varphi_i^0}\; \mathrm{Im} (y_\ell^{\varphi^0_i})
\; \hat{C}^{\varphi_i^0}\, .
\end{equation}

The contributions from diagrams \ref{fig:hiA2HDM}.4, \ref{fig:hiA2HDM}.7, \ref{fig:hiA2HDM}.8 and \ref{fig:hiA2HDM}.14 are proportional to the light-quark mass $m_q$ and, therefore, vanish in our massless approximation. Diagrams \ref{fig:hiA2HDM}.1, \ref{fig:hiA2HDM}.3, \ref{fig:hiA2HDM}.11 and \ref{fig:hiA2HDM}.13 in Feynman gauge and diagrams \ref{fig:hiA2HDM}.1, \ref{fig:hiA2HDM}.3, \ref{fig:hiA2HDM}.5, \ref{fig:hiA2HDM}.6, \ref{fig:hiA2HDM}.9 and \ref{fig:hiA2HDM}.10 in unitary gauge generate a divergent contribution, which is not eliminated by the GIM mechanism; \textit{i.e.}, it remains even after summing over contributions of the three virtual up-type quarks. This divergence matches exactly the expected behaviour predicted through the RGEs, which originated in the local term $\cL_{\rm FCNC}$. Thus, the one-loop divergence is cancelled by the renormalization of the coupling $\cC$ in Eq.~\eqn{eq:Crenorm} which, moreover, reabsorbs the $\mu$ dependence of the loops into the combination $\cC_R(M_W) = \cC_R(\mu) -\ln{(M_W/\mu)}$.

The scalar penguin diagrams \ref{fig:hiA2HDM}.2, \ref{fig:hiA2HDM}.12, \ref{fig:hiA2HDM}.15 and \ref{fig:hiA2HDM}.16 involve the cubic couplings $\varphi_i^0H^+H^-$, $\varphi_i^0G^+G^-$, $\varphi_i^0H^+G^-$ and $\varphi_i^0G^+H^-$, respectively, which are functions of the scalar-potential parameters. Since the last three couplings can be fully determined in terms of the vacuum expectation value $v$ and the scalar masses and mixings, we can express the total scalar-exchange (tree-level plus one-loop penguin) contribution in the form:
\beqn \label{eq:scalarPenguinRes}
\hat{C}^{\varphi_i^0} & = & x_t\;\Biggl\{
\frac{1}{2 x_{\varphi^0_i}}\; (\varsigma_u-\varsigma_d)\, (1+\varsigma^*_u\varsigma_d)\, (\cR_{i2} + i \cR_{i3})
\; \cC_R(M_W)
+ \frac{v^2}{M^2_{\varphi^0_i}}\; \lambda^{\varphi_i^0}_{H^+H^-}\;
g_0^{\phantom{()}}(x_t,x_{H^+},\varsigma_u,\varsigma_d)
\no\\
&&\hskip .5cm \mbox{} + \sum_{j=1}^3\;\cR_{ij}\,\xi_j\; \left[ \frac{1}{2 x_{\varphi^0_i}}\; g_j^{(a)}(x_t,x_{H^+},\varsigma_u,\varsigma_d) + g_j^{(b)}(x_t,x_{H^+},\varsigma_u,\varsigma_d)\right] \Biggr\}\, ,
\eeqn
where
$\lambda^{\varphi_i^0}_{H^+H^-} = \lambda_3\cR_{i1} + \lambda_7^R\cR_{i2}-\lambda_7^I\cR_{i3}$, $\xi_1=\xi_2 = 1$ and $\xi_3=i$.
The functions $g_0^{\phantom{()}}(x_t,x_{H^+},\varsigma_u,\varsigma_d)$, $g_j^{(a)}(x_t,x_{H^+},\varsigma_u,\varsigma_d)$ and $g_j^{(b)}(x_t,x_{H^+},\varsigma_u,\varsigma_d)$ are given in the appendix, both in the Feynman and unitary gauges, together with the separate contributions from each diagram in Fig.~\ref{fig:hiA2HDM}. In the limit $\varsigma_{u,d}\to 0$, $x_{H,A}\to\infty$, $x_h\to x_{h_{\rm SM}}$, $\cR_{i2,i3}\to 0$, $\cR_{11}\to1$, this result reduces to the SM expression in Eqs.~\eqn{eq:csSM1} and \eqn{eq:csSM2}.

The orthogonality relation~\cite{A2HDM:collider2}
\bel{eq:ortho}
\sum_{i=1}^3\; y_\ell^{\varphi_i^0}\;\cR_{ij}\; =\;
\delta_{j1} + \left( \delta_{j2} + i\,\delta_{j3} \right)\,\varsigma_\ell
\ee
allows us to separate the total contribution from the functions $g_j^{(b)}(x_t,x_{H^+},\varsigma_u,\varsigma_d)$, which does not depend on the neutral scalar masses:
\beqn \label{eq:gb-contrib1}
\left. C_S^{\varphi_i^0,\, \rm A2HDM}\right|_{g^{(b)}} & = &
x_t\;\left[g_1^{(b)} + \mathrm{Re}(\varsigma_\ell) \; g_2^{(b)}
- i\,\mathrm{Im}(\varsigma_\ell) \; g_3^{(b)} \right]\, ,\\[0.2cm]
       \label{eq:gb-contrib2}
\left. C_P^{\varphi_i^0,\, \rm A2HDM}\right|_{g^{(b)}} & = &
x_t\;\left[i\,\mathrm{Im}(\varsigma_\ell)\; g_2^{(b)} - \mathrm{Re}(\varsigma_\ell)\; g_3^{(b)} \right]\, .
\eeqn
It is also noted that the functions $g_j^{(b)}(x_t,x_{H^+},\varsigma_u,\varsigma_d)$ only receive contributions in the Feynman gauge, because they arise from the scalar penguin diagrams involving the Goldstone bosons. Actually, the gauge dependent pieces from the box diagrams shown in Figs.~\ref{fig:BoxSM} and \ref{fig:BoxA2HDM} are exactly cancelled by these terms:
\begin{align} \label{eq:CSSM-gauge}
C^{\rm box,\, \rm SM}_{S, \,\rm Unitary} \; - \;  C^{\rm box,\, \rm SM}_{S, \,\rm Feynman}  & \; = \; x_t\, g_1^{(b)} \, , \\[0.2cm]
      \label{eq:CSA2HDM-gauge}
C^{\rm box,\, \rm A2HDM}_{S, \,\rm Unitary} \; - \;  C^{\rm box,\, \rm A2HDM}_{S, \,\rm Feynman} & \; = \; x_t\left[\mathrm{Re}(\varsigma_\ell) \; g_2^{(b)} - i\,\mathrm{Im}(\varsigma_\ell) \; g_3^{(b)}\right] \, ,\\[0.2cm]
       \label{eq:CPA2HDM-gauge}
C^{\rm box,\, \rm A2HDM}_{P, \,\rm Unitary} \; - \;  C^{\rm box,\, \rm A2HDM}_{P, \,\rm Feynman} & \; = \; x_t\left[ i\,\mathrm{Im}(\varsigma_\ell)\; g_2^{(b)} - \mathrm{Re}(\varsigma_\ell)\; g_3^{(b)} \right]\, .
\end{align}

The remaining contributions in Eq.~(\ref{eq:scalarPenguinRes}), which are all proportional to $1/M_{\varphi_i^0}^2$, are gauge independent but are sensitive to the scalar mixing parameters. Nevertheless, a naive mixing-independent estimate can be obtained in the limit of degenerate neutral-scalar masses:
\beqn \label{eq:CS-contrib}
\left. C_S^{\varphi_i^0,\, \rm A2HDM}\right|_{\cC + g_0^{\phantom{()}} + g^{(a)}}^{x_h = x_H = x_A} & = &
\frac{x_t}{2 x_h}\;\Biggl\{
 (\varsigma_u-\varsigma_d)\,(1+\varsigma^*_u\varsigma_d)\, \cC_R(M_W)\,
\Bigl[\mathrm{Re}(\varsigma_\ell)-i\,\mathrm{Im}(\varsigma_\ell)\Bigr]
\no\\ &&\hskip .75cm\mbox{}
+\frac{2 v^2}{M_W^2}\; g_0^{\phantom{()}}\; \Bigl[\lambda_3 + \lambda_7^R\,\mathrm{Re}(\varsigma_\ell)+ \lambda_7^I\, \mathrm{Im}(\varsigma_\ell)\Bigr]
\no\\ &&\hskip .75cm\mbox{}
+ g_1^{(a)} 
+ \mathrm{Re}(\varsigma_\ell) \; g_2^{(a)} 
- i\,\mathrm{Im}(\varsigma_\ell) \; g_3^{(a)} 
\Biggr\}\, , \\[0.2cm]
       \label{eq:CP-contrib}
\left. C_P^{\varphi_i^0,\, \rm A2HDM}\right|_{\cC + g_0^{\phantom{()}} + g^{(a)}}^{x_h = x_H = x_A} & = &
\frac{x_t}{2 x_h}\;\Biggl\{
 (\varsigma_u-\varsigma_d)\,(1+\varsigma^*_u\varsigma_d)\, \cC_R(M_W)\,
\Bigl[i\,\mathrm{Im}(\varsigma_\ell)-\mathrm{Re}(\varsigma_\ell)\Bigr]
\no\\ &&\hskip .75cm\mbox{}
+\frac{2 v^2}{M_W^2}\; g_0^{\phantom{()}}\; i\,\Bigl[\lambda_7^R\, \mathrm{Im}(\varsigma_\ell)- \lambda_7^I\, \mathrm{Re}(\varsigma_\ell)\Bigr]
\no\\ &&\hskip .75cm\mbox{}
+
i\,\mathrm{Im}(\varsigma_\ell)\; g_2^{(a)} 
- \mathrm{Re}(\varsigma_\ell)\; g_3^{(a)} 
\Biggr\}\, .
\eeqn
We shall perform our phenomenological analyses in the CP-conserving limit, with real potential and alignment parameters, where $A=S_3$ is a CP-odd state while $H$ and $h$ are two CP-even states defined by the rotation in Eq.~(\ref{eq:CPC_mixing}). The $1/x_{\varphi^0_i}$ contributions take then the form:
\beqn
\left. C_S^{\varphi_i^0,\, \rm A2HDM}\right|_{\cC + g_0^{\phantom{()}} + g^{(a)}}^{\rm CP\,\mathrm{con.}} & = &
\frac{x_t}{2 x_h}\,\left( c_{\tilde{\alpha}} + s_{\tilde{\alpha}}\,\varsigma_\ell\right)\,
\Biggl\{s_{\tilde{\alpha}}\, (\varsigma_u-\varsigma_d)\, (1+\varsigma_u\,\varsigma_d)\, \cC_R(M_W)
\no\\ &&\hskip 3.0cm\mbox{}
+ \left( c_{\tilde{\alpha}}\, \lambda_3 + s_{\tilde{\alpha}}\, \lambda_7 \right)\,
\frac{2 v^2}{M_W^2}\; g_0^{\phantom{()}}+ c_{\tilde{\alpha}}\; g_1^{(a)} + s_{\tilde{\alpha}}\; g_2^{(a)}\Biggr\}
\no\\ &+&
\frac{x_t}{2 x_{H}}\,\left( c_{\tilde{\alpha}}\,\varsigma_\ell -s_{\tilde{\alpha}}\right)\,
\Biggl\{ c_{\tilde{\alpha}}\,  (\varsigma_u-\varsigma_d)\, (1+\varsigma_u\,\varsigma_d)\, \cC_R(M_W)
\label{eq:SexchCPC-S}\\ &&\hskip 3.0cm\mbox{}
- \left( s_{\tilde{\alpha}}\, \lambda_3 - c_{\tilde{\alpha}}\, \lambda_7 \right)\,
\frac{2 v^2}{M_W^2}\; g_0^{\phantom{()}}- s_{\tilde{\alpha}}\; g_1^{(a)} + c_{\tilde{\alpha}}\; g_2^{(a)}\Biggr\}\, ,\no
\\[0.2cm]
\left. C_P^{\varphi_i^0,\, \rm A2HDM}\right|_{\cC + g_0^{\phantom{()}} + g^{(a)}}^{\rm CP\,\mathrm{con.}} & = &
-\varsigma_\ell\;\frac{x_t}{2 x_A}\;
\left[ (\varsigma_u-\varsigma_d)\,(1+\varsigma_u\,\varsigma_d)\, \cC_R(M_W) + g_3^{(a)}\right]\, ,
\label{eq:SexchCPC-P}
\eeqn
where $c_{\tilde{\alpha}}=\cos\tilde{\alpha}$ and $s_{\tilde{\alpha}}=\sin\tilde{\alpha}$. For degenerate neutral scalars, this reproduces the results in Eqs.~\eqn{eq:CS-contrib} and \eqn{eq:CP-contrib} (in the CP-conserving limit).

The terms proportional to $C_R(M_W)$ in Eqs.~(\ref{eq:SexchCPC-S}) and (\ref{eq:SexchCPC-P}) are absent in $\mathcal{Z}_2$-symmetric models, because the alignment conditions are protected by the $\mathcal{Z}_2$ symmetry at any scale. In the particular case of the type-II 2HDM at large $\tan\beta$, the only terms enhanced by a factor $\tan^2\beta$ originate from the $\varsigma_\ell g_2^{(a)}$~(for $C_S$) and $\varsigma_\ell g_3^{(a)}$~(for $C_P$) contributions, due to the factors $\varsigma_d^2\varsigma_u^{\ast}$ and $\varsigma_d$ in the definitions for $g_{2}^{(a)}$ and $g_{3}^{(a)}$~(see Eqs.~(\ref{eq:g2a_def}) and (\ref{eq:g3a_def})). In this specific case, our results agree with the ones calculated in Ref.~\cite{Logan:2000iv}. Especially, we confirmed the observation that the dependence on the masses of the neutral Higgs bosons from the penguin and fermion self-energy diagrams drops out in their sum without invoking any relation between the mixing angle and the Higgs masses~\cite{Logan:2000iv}.

\subsection{$\boldsymbol{\Bqll}$ branching ratio}
\label{sec:BR}

Due to the pseudoscalar nature of the $B_q$ meson, only the following two hadronic matrix elements are involved in $\Bqll$ decays:
\begin{align} \label{eq:decayconstant_def}
 \left\langle 0 |\bar{q}\, \gamma_\mu \gamma_5\, b | \bar{B}_q(p) \right \rangle & \; = \; i f_{B_q} p_\mu\,,\no\\
 \left\langle 0 |\bar{q}\, \gamma_5\, b | \bar{B}_q(p) \right \rangle & \; = \; -i f_{B_q} \frac{M_{B_q}^2}{m_b+m_q}\,,
\end{align}
where $f_{B_q}$ and $M_{B_q}$ are the $B_q$-meson decay constant and mass, respectively. The second equation follows from the first one by using the QCD equation of motion for the quark fields.

Starting with Eq.~(\ref{eq:Heff}) and using Eq.~(\ref{eq:decayconstant_def}), we can express the branching ratio of $\Bqll$ decays as
\begin{eqnarray} \label{eq:BR}
 \mathcal{B}(B_q^0\to \ell^+\ell^-) & = & \frac{\tau_{B_q}\,G_F^4\,M_W^4}{8\pi^5}\, \left|V_{tb}\,V_{tq}^*\,C_{10}^{\rm SM}\right|^2\, f_{B_q}^2 M_{B_q} m_{\ell}^2\, \sqrt{1-\frac{4m_{\ell}^2}{M_{B_q}^2}}\; \Big[\,|P|^2+|S|^2\,\Big]\,,\no\\[0.2cm]
 & = & \mathcal{B}(B_q^0\to \ell^+\ell^-)_{\rm SM}\,\Big[\,|P|^2+|S|^2\,\Big]\,,
\end{eqnarray}
where $\tau_{B_q}$ is the $B_q$-meson mean lifetime, and $P$ and $S$ are defined, respectively, as~\cite{Buras:2013uqa,DeBruyn:2012wk}
\begin{align}
 P &\equiv\, \frac{C_{10}}{C^{\rm SM}_{10}} + \frac{M^2_{B_q}}{2M^2_W} \left(\frac{m_b}{m_b+m_q}\right)\,\frac{C_P-C_P^{\mathrm{SM}}}{C^{\rm SM}_{10}} \,\equiv\, |P|\; e^{i\phi_P}\,,
 \label{eq:P}\\[0.2cm]
 S &\equiv\, \sqrt{1-\frac{4m^2_\ell}{M^2_{B_q}}}\; \frac{M^2_{B_q}}{2M^2_W} \left(\frac{m_b}{m_b+m_q}\right)\,\frac{C_S-C_S^{\mathrm{SM}}}{C^{\rm SM}_{10}} \,\equiv\, |S|\; e^{i\phi_S}\,.
 \label{eq:S}
\end{align}
We have approximated the negligibly small~(and usually neglected) SM
scalar/pseudoscalar contributions\footnote{Here, $C_S^{\mathrm{SM}}$ and $C_P^{\mathrm{SM}}$ denote the full SM contribution, including the Higgs-penguin terms.} $C_S^{\mathrm{SM}}$ and $C_P^{\mathrm{SM}}$ to first order in $M^2_{B_q}/M^2_W$. In the SM, $P=1$ and $S=0$. In a generic case, however, $P$ and $S$ can carry nontrivial CP-violating phases $\phi_P$ and $\phi_S$. It is also noted that, even in models with comparable Wilson coefficients, the contributions from ${\cal O}_{S}$ and ${\cal O}_{P}$ are suppressed by a factor $M_{B_q}^2/M_W^2$ with respect to that from ${\cal O}_{10}$. Therefore, unless there were large enhancements for $C_S$ and $C_P$, the coefficient $C_{10}$ still provides the dominant contribution to the branching ratio.

In order to compare with the experimental measurement, the effect of $B^0_q-\bar B^0_q$ oscillations should be taken into account, and the resulting averaged time-integrated branching ratio is given by~\cite{Buras:2013uqa,DeBruyn:2012wk}
\begin{equation} \label{eq:BR_bar}
 \overline{\mathcal{B}}(B_q^0\to \ell^+\ell^-)\, =\, \bigg[ \frac{1+\mathcal{A}_{\Delta\Gamma}^{\ell\ell}\,y_q}{1-y_q^2}\bigg]\,\mathcal{B}(B_q^0\to \ell^+\ell^-)\,,
\end{equation}
where $\mathcal{A}_{\Delta\Gamma}^{\ell\ell}$ is a time-dependent observable introduced firstly in Ref.~\cite{DeBruyn:2012wk}, and $y_q$ is related to the decay width difference $\Delta\Gamma_q$ between the two $B_q$-meson mass eigenstates,
\begin{equation}
 y_q \,\equiv\, \frac{\Gamma^{q}_L-\Gamma^{q}_H}{\Gamma^{q}_L+\Gamma^{q}_H}\, =\, \frac{\Delta\Gamma_q}{2\Gamma_q}\,,
\end{equation}
with $\Gamma^{q}_{H(L)}$ denoting the heavier~(lighter) eigenstate decay width and $\Gamma_q=\tau_{B_q}^{-1}$ the average $B_q$-meson width. Within the SM, $\mathcal{A}_{\Delta\Gamma}^{\ell\ell}=1$ and the averaged time-integrated branching ratio is given by
\begin{eqnarray} \label{eq:BR_bar_SM}
 \overline{\mathcal{B}}(B_q^0\to \ell^+\ell^-)_{\rm SM} & = & \frac{1}{1-y_q}\; \mathcal{B}(B_q^0\to \ell^+\ell^-)_{\rm SM}\,,\no\\[0.2cm]
 & = & \frac{G_F^4\,M_W^4}{8\pi^5\,\Gamma^{q}_H}\, \left|V_{tb}\,V_{tq}^*\,C_{10}^{\rm SM}\right|^2\, f_{B_q}^2 M_{B_q} m_{\ell}^2\, \sqrt{1-\frac{4m_{\ell}^2}{M_{B_q}^2}}\,.
\end{eqnarray}
By exploiting Eqs.~(\ref{eq:BR}) and (\ref{eq:BR_bar_SM}), we can rewrite Eq.~(\ref{eq:BR_bar}) as
\begin{eqnarray} \label{eq:BR_bar_CPcons}
 \overline{\mathcal{B}}(B_q^0\to \ell^+\ell^-) & = & \bigg[ \frac{1+\mathcal{A}_{\Delta\Gamma}^{\ell\ell}\,y_q}{1+y_q}\bigg]\, \Big[|P|^2+|S|^2\Big]\, \overline{\mathcal{B}}(B_q^0\to \ell^+\ell^-)_{\rm SM}\, \,,\no\\[0.2cm]
 & \dot= & \overline{\mathcal{B}}(B_q^0\to \ell^+\ell^-)_{\rm SM}\,\bigg[|P|^2+\Big(1-\frac{\Delta\Gamma_q}{\Gamma^q_L}\Big)|S|^2\bigg]\,,
\end{eqnarray}
where the second line is valid only in the absence of beyond-SM sources of CP violation,
which will be assumed in the following.\footnote{The explicit formulae in a generic case with new CP-violating phases could be found in Refs.~\cite{Hermann:2013kca,Buras:2013uqa,DeBruyn:2012wk}.}

\section{Numerical results}
\label{sec:numberical}

\subsection{Input parameters}
\label{sec:inputs}

\begin{table}[t]
\centering
\tabcolsep 0.20in
\begin{tabular}{|l|l|}
\hline & \\[-0.6cm]
  $G_F = 1.1663787 \times 10^{-5}~\mathrm{GeV}^{-2}$  \hfill\cite{Beringer:1900zz}  &
  $f_{B_s} = 227.7 \pm 4.5~\mathrm{MeV}$                  \hfill\cite{Aoki:2013ldr} \\
  $\alpha_s(M_Z) = 0.1185 \pm 0.0006$                 \hfill\cite{Beringer:1900zz}  &
  $f_{B_d} = 190.5 \pm 4.2~\mathrm{MeV}$                  \hfill\cite{Aoki:2013ldr} \\
  $\Delta\alpha_{\rm hadr}(M_Z) = 0.02772 \pm 0.00010$ \hfill\cite{Beringer:1900zz} &
  $\tau_{B_s} = 1.516 \pm 0.011~\mathrm{ps}$           \hfill\cite{Amhis:2012bh}    \\
  $M_Z = 91.1876 \pm 0.0021~\mathrm{GeV}$              \hfill\cite{Beringer:1900zz} &
  $\tau_{B_d} = 1.519 \pm 0.007~\mathrm{ps}$           \hfill\cite{Amhis:2012bh}    \\
  $M_t = 173.34\pm0.27\pm1.71~\mathrm{GeV}$                 \hfill\cite{ATLAS:2014wva}   &
  $1/\Gamma_H^s = 1.615 \pm 0.021~\mathrm{ps}$         \hfill\cite{Amhis:2012bh}    \\
  $M_{h_{\rm SM}} = 125.9 \pm0.4~\mathrm{GeV}$         \hfill\cite{Beringer:1900zz} &                $1/\Gamma_L^s = 1.428 \pm 0.013~\mathrm{ps}$         \hfill\cite{Amhis:2012bh}    \\
  $M_{B_s} = 5366.77 \pm 0.24~\mathrm{MeV}$            \hfill\cite{Beringer:1900zz} &
  $\Delta\Gamma_s = 0.081 \pm 0.011~\mathrm{ps}^{-1}$  \hfill\cite{Amhis:2012bh}    \\
  $M_{B_d} = 5279.58 \pm 0.17~\mathrm{MeV}$            \hfill\cite{Beringer:1900zz} &
  $|V_{cb}| = (42.42\pm 0.86) \times 10^{-3}$             \hfill\cite{Gambino:2013rza} \\
  $m_b(m_b) = 4.18 \pm 0.03~\mathrm{GeV}$              \hfill\cite{Beringer:1900zz} &
  $|V^{\ast}_{tb}V^{\phantom{*}}_{ts}/V_{cb}| = 0.980 \pm 0.001$
  \hfill\cite{Charles:2004jd,Ciuchini:2000de} \\
  $m_s(2~\mathrm{GeV}) = 95 \pm 5~\mathrm{MeV}$        \hfill\cite{Beringer:1900zz} & $|V^{\ast}_{tb}V^{\phantom{*}}_{td}| = 0.0088\pm0.0003$ \hfill\cite{Charles:2004jd,Ciuchini:2000de}  \\
  $m_{\mu} = 105.65837~\mathrm{MeV}$            \hfill\cite{Beringer:1900zz} & \\
[0.15cm]
\hline
\end{tabular}
\caption{\small Relevant input parameters used in our numerical analysis.}
\label{tab:input}
\end{table}

To evaluate numerically the branching ratios in Eqs.~(\ref{eq:BR_bar_SM}) and (\ref{eq:BR_bar_CPcons}), we need several input parameters collected in Table~\ref{tab:input}. For the matching scale $\mu_0\sim \mathcal{O}(M_{W})$ and the low-energy scale $\mu_b\sim \mathcal{O}(m_b)$, we fix them to $\mu_0=160~\mathrm{GeV}$ and $\mu_b=5~\mathrm{GeV}$~\cite{Bobeth:2013uxa}. In addition, the on-shell scheme is adopted for the EW parameters, which means that the $Z$-boson and top-quark masses coincide with their pole masses, and the weak angle is given by $s_W^2 \equiv 1-M_W^2/M_Z^2$, where $M_W=80.359\pm0.012~\mathrm{GeV}$ is the $W$-boson on-shell mass obtained according to the fit formulae in Eqs.~(6) and (9) of Ref.~\cite{Awramik:2003rn}.

For the top-quark mass, we assume that the combined measurement of Tevatron and LHC~\cite{ATLAS:2014wva} corresponds to the pole mass, but increase its systematic error by
$1~\mathrm{GeV}$ to account for the intrinsic ambiguity in the $m_t$ definition; \textit{i.e.} we shall take $M_t=(173.34\pm0.27\pm1.71)~\mathrm{GeV}$. With the aid of the \texttt{Mathematica} package \texttt{RunDec}~\cite{Chetyrkin:2000yt}, four-loop QCD RGEs are applied to evolve the strong coupling $\alpha_s(\mu)$ as well as the $\overline{\rm MS}$ renormalized masses $m_t(\mu)$ and $m_{b,s}(\mu)$ between different scales, and a three-loop relation has been used to convert the pole mass $M_t$ to the scale-invariant mass $m_t(m_t)$, which gives $m_t(m_t)\simeq163.30~\mathrm{GeV}$.

The decay constants $f_{B_q}$ are taken from the updated FLAG~\cite{Aoki:2013ldr} average of $N_f=2+1$ lattice determinations, which are consistent with the naive weighted average of $N_f=2+1$~\cite{McNeile:2011ng,Bazavov:2011aa,Christ:2014uea} and $N_f=2+1+1$~\cite{Dowdall:2013tga,Carrasco:2013naa} results. For the $B_q$-meson lifetimes, while a sizable decay width difference $\Delta\Gamma_s$ has been established~\cite{Amhis:2012bh}, the approximation $1/\Gamma_H^d\simeq1/\Gamma_L^d\equiv\tau_{B_d}$ can be safely set, given the tiny SM expectation for $\Delta\Gamma_d/\Gamma_d$~\cite{Lenz:2011ti}.

For the CKM matrix element $|V_{cb}|$, we adopt the recent inclusive fit performed by taking into account both the semileptonic data and the precise quark mass determinations from flavour-conserving processes~\cite{Gambino:2013rza}. However, one should be aware
of the present disagreement between inclusive and exclusive determinations~\cite{Aoki:2013ldr}. With $|V_{cb}|$ fixed in this way, the needed CKM factors are then obtained~(within the SM) from the accurately known ratio $|V^{\ast}_{tb}V^{\phantom{*}}_{ts}/V_{cb}|$~\cite{Charles:2004jd,Ciuchini:2000de}.

\subsection{SM predictions}
\label{sec:smresults}

Within the SM, only the Wilson coefficient $C_{10}^{\rm SM}$ is relevant and, using the fitting formula in Eq.~(4) of Ref.~\cite{Bobeth:2013uxa} (which has been transformed to our convention for the effective Hamiltonian),
\begin{align} \label{eq:C10SMfit}
C^{\rm SM}_{10} &= -0.9604 \left[\frac{M_t}{173.1~\mathrm{GeV}}\right]^{1.52} \left[\frac{\alpha_s(M_Z)}{0.1184}\right]^{-0.09} + 0.0224 \left[\frac{M_t}{173.1~\mathrm{GeV}}\right]^{0.89}\, \left[\frac{\alpha_s(M_Z)}{0.1184}\right]^{-0.09}\,,\no \\[0.2cm]
&= -0.9380 \left[\frac{M_t}{173.1~\mathrm{GeV}}\right]^{1.53} \left[\frac{\alpha_s(M_Z)}{0.1184}\right]^{-0.09}\,.
\end{align}
The EW and QCD factors introduced in Eq.~(\ref{eq:C10SM}) are extracted as:
\begin{equation} \label{eq:SMcorrection}
 \eta_Y^{\rm EW} = 0.977\,, \qquad \qquad \eta_Y^{\rm QCD} = 1.010\,.
\end{equation}

With the input parameters collected in Table~\ref{tab:input}, the SM predictions for the branching ratios of $\Bqll$ decays are:
\begin{align} \label{eq:BqllSM}
 \overline{\mathcal{B}}(B_s \to e^+ e^-) & = (8.58 \pm 0.59) \times 10^{-14}\,, \no \\[0.2cm]
 \overline{\mathcal{B}}(B_s \to \mu^+ \mu^-) & = (3.67 \pm 0.25) \times 10^{-9}\,, \no \\[0.2cm]
 \overline{\mathcal{B}}(B_s \to \tau^+ \tau^-) & = (7.77 \pm 0.53) \times 10^{-7}\,, \no \\[0.2cm]
 \overline{\mathcal{B}}(B_d \to e^+ e^-) & = (2.49 \pm 0.22) \times 10^{-15}\,, \no \\[0.2cm]
 \overline{\mathcal{B}}(B_d \to \mu^+ \mu^-) & = (1.06 \pm 0.10) \times 10^{-10}\,, \no \\[0.2cm]
 \overline{\mathcal{B}}(B_d \to \tau^+ \tau^-) & = (2.23 \pm 0.20) \times 10^{-8}\,,
\end{align}
where a $1.5\%$ nonparametric uncertainty has been set to the branching ratios, and the main parametric uncertainties come from $f_{B_q}$ and the CKM matrix elements~\cite{Bobeth:2013uxa}. The small differences with respect to the results given in Ref.~\cite{Bobeth:2013uxa} are due to our slightly different (more conservative) input value for the top-quark mass $M_t$.

In order to explore constraints on the model parameters, it is convenient to introduce the ratio~\cite{Buras:2013uqa,DeBruyn:2012wk}
\begin{equation} \label{eq:R_bar_CPcons}
 \overline{R}_{q\ell}  \,\equiv\,  \frac{\overline{\mathcal{B}}(B_q^0\to \ell^+\ell^-)}{\overline{\mathcal{B}}(B_q^0\to \ell^+\ell^-)_{\rm SM}}
\, =\, \bigg[\,|P|^2+\Big(1-\frac{\Delta\Gamma_q}{\Gamma^q_L}\,\Big)|S|^2\bigg]\,,
\end{equation}
where the hadronic factors and CKM matrix elements cancel out. Combining the theoretical SM predictions in Eq.~(\ref{eq:BqllSM}) with the experimental results in Eqs.~(\ref{eq:Bsmmexp}) and (\ref{eq:Bdmmexp}), we get
\begin{equation} \label{eq:R_bar_results}
 \overline{R}_{s\mu} = 0.79 \pm 0.20\,, \qquad \qquad \overline{R}_{d\mu}  = 3.38^{+1.53}_{-1.35}\,,
\end{equation}
to be compared with the SM expectation $\overline{R}^{\rm SM}_{s\mu}=\overline{R}^{\rm SM}_{d\mu}=1$.

Since only the $B_s\to \mu^+\mu^-$ branching ratio is currently measured with a signal significance of $\sim4.0\sigma$~\cite{CMSandLHCbCollaborations:2013pla}, we shall investigate the allowed parameter space of the A2HDM under the constraint from $\overline{R}_{s\mu}$ given in Eq.~(\ref{eq:R_bar_results}). Although the experimental uncertainty is still quite large, it has already started to put stringent constraints on many models beyond the SM~\cite{Buras:2013uqa}.

Notice that, in addition to modifying the ratios $\overline{R}_{q\ell}$, the scalar contributions to $B^0_q$--$\bar B^0_q$ mixings also change the fitted values of the relevant CKM parameters and, therefore, the normalization $\overline{\mathcal{B}}(B_q^0\to \ell^+\ell^-)_{\rm SM}$. This should be taken into account, once more precise $B_q^0\to \ell^+\ell^-$ data becomes available, through a combined global fit.

\subsection{Results in the A2HDM}
\label{sec:A2HDMresult}

\subsubsection{Choice of model parameters}

In the following we assume that the Lagrangian of the scalar sector preserves the CP symmetry \textit{i.e.}, that the only source of CP violation is still due to the CKM matrix. This makes all the alignment and scalar-potential parameters real. Assuming further that the lightest CP-even scalar $h$ corresponds to the observed neutral boson with $M_h\simeq126~\mathrm{GeV}$, there are ten free parameters in our calculation: three alignment parameters $\varsigma_f$, three scalar masses~($M_H$, $M_A$, $M_{H^\pm}$), one mixing angle $\tilde\alpha$, two scalar-potential couplings~($\lambda_3$, $\lambda_7$), and the misalignment parameter $C_R(M_W)$.

In order to gain insight into the parameter space allowed by $\Bqll$ decays, it is necessary to take into account information about the $h(126)$ collider data and flavour physics constraints, as well as EW precision observables, which will be crucial for making simplifying assumptions and reducing the number of relevant variables. Explicitly, the following constraints and assumptions on the model parameters are taken into account:
\begin{itemize}
  \item Firstly, the mixing angle $\tilde\alpha$ is constrained at $|\cos\tilde\alpha|>0.90$~($68\%$~CL) through a global fit to the latest LHC and Tevatron data for the $h(126)$ boson~\cite{A2HDM:collider2}, which is very close to the SM limit; \textit{i.e.}, the lightest CP-even scalar $h$ behaves like the SM Higgs boson.

  \item To assure the validity of perturbative unitarity in the scalar-scalar scattering amplitudes, upper bounds on the quartic Higgs self-couplings are usually imposed by requiring them to be smaller than $8\pi$~\cite{2HDM:review}; \textit{i.e.}, $|\lambda_{3,7}|\lesssim 8\pi$.

  \item With our convention, the lower bound on the heavier CP-even scalar mass is $M_H\geq M_h\simeq126~\mathrm{GeV}$. Much lower values of $M_A$ are still allowed experimentally. There are, however, no stringent upper limits on these masses. Here we limit them at $M_H \in [130, 500]~\mathrm{GeV}$ and $M_A \in [80,500]~\mathrm{GeV}$.

  \item  The charged Higgs mass is assumed to lie in the range $M_{H^{\pm}} \in[80,500]~\mathrm{GeV}$, which would require $|\varsigma_u|\leq 2$ to be compatible with the present data on loop-induced processes, such as $Z\to \bar{b}b$, $b\to s \gamma$ and $B_{s,d}^0-\bar{B}_{s,d}^0$ mixing, as well as the $h(126)$ decays~\cite{A2HDM:flavour,A2HDM:collider2}.

  \item The alignment parameters $\varsigma_d$ and $\varsigma_\ell$ are only mildly constrained through phenomenological requirements involving other model parameters. As in our previous works, we restrict them at $|\varsigma_{d,\ell}|\leq50$~\cite{A2HDM:flavour}.

  \item At present, there are no useful constraints on the misalignment parameter $C_R(M_W)$. For simplicity, it is assumed to be zero.
\end{itemize}

Numerically, it is found that the ratio $\overline{R}_{s\mu}$ is less sensitive to the scalar-potential couplings $\lambda_3$ and $\lambda_7$ than to the other model parameters, especially when the alignment parameters are small and/or the neutral scalar masses are large. The mixing angle $\tilde\alpha$, when constrained in the range $\cos\tilde\alpha\in[0.9, 1]$, is also found to have only a marginal impact on $\overline{R}_{s\mu}$. Thus, for simplicity, we shall assign the following values to these parameters:
\begin{equation} \label{eq:value_l3l7ca}
\lambda_3 =  \lambda_7 = 1, \qquad \qquad \cos\tilde\alpha = 0.95\,.
\end{equation}

As can be seen from Eqs.~(\ref{eq:P}) and (\ref{eq:S}), the Wilson coefficients $C_S$ and $C_P$ are always accompanied with the power-suppressed factor $M^2_{B_q}/M^2_W$ compared to $C_{10}$. The NP contribution to $C_{10}$ is, however, proportional to $|\varsigma_u|^2$ and depends only on the charged-scalar mass. It is, therefore, interesting to discuss the following two special cases with respect to the choice of the alignment parameters: The first one is when $|\varsigma_{d,\ell}|\lesssim |\varsigma_u|\leq 2$, where the NP contribution is dominated by $C_{10}$ while $C_S$ and $C_P$ are negligible. The second one is when $|\varsigma_{d,\ell}| \gg |\varsigma_u|$, which means that $C_S$ and $C_P$ play a significant role.

\subsubsection{Small $\boldsymbol{\varsigma_{d,\ell}}$}

When the alignment parameters $\varsigma_{d,\ell}$ are of the same size as (or smaller than) $\varsigma_u$, the NP contributions from $C_S$ and $C_P$ are negligible. In this case, we need only to focus on the Wilson coefficient $C_{10}$, which is the sum of the SM contribution $C^{\rm SM}_{10}$ and the charged-Higgs contribution $C^{\rm A2HDM}_{10}$ due to $Z$-penguin diagrams shown in Fig.~\ref{fig:ZA2HDM}. The latter involves only two free parameters, $\varsigma_u$ and $M_{H^\pm}$, and goes to zero when $\varsigma_u \to 0$ and/or $M_{H^\pm}\to\infty$.

\begin{figure}[t]
  \centering
  \includegraphics[width=\textwidth]{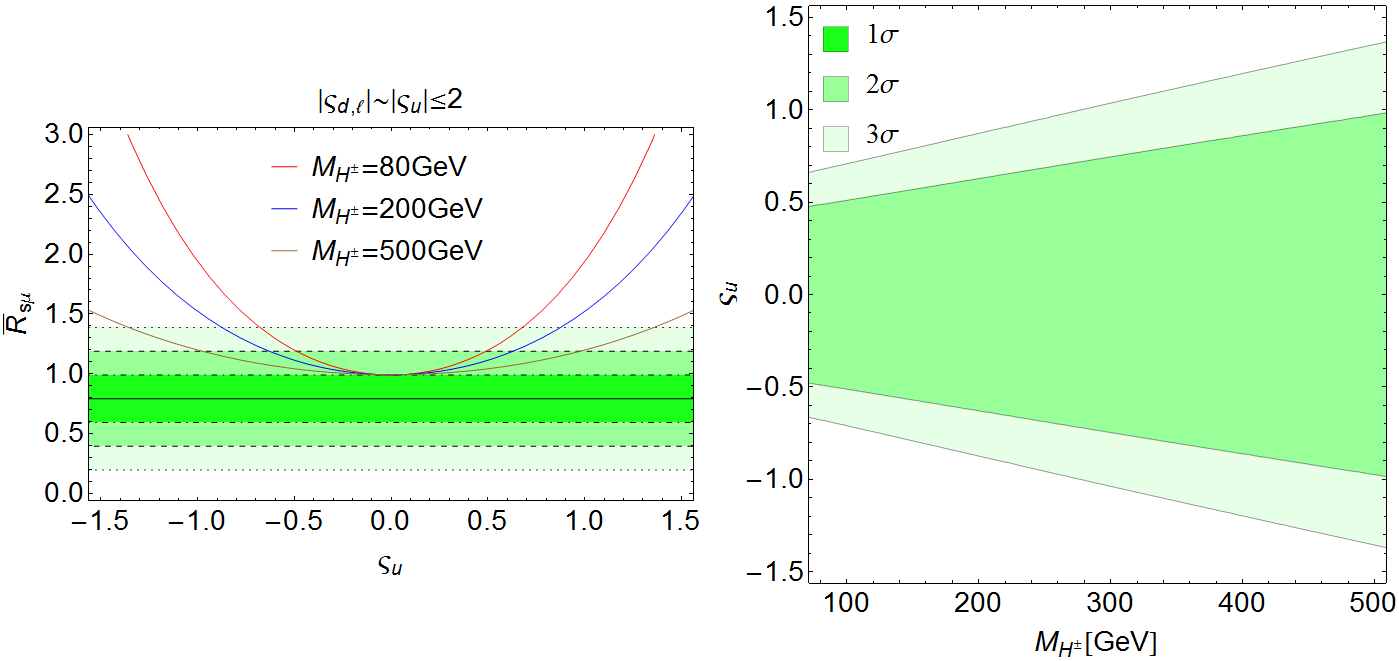}
  \caption{\small Dependence of $\bar R_{s\mu}$ on $\varsigma_u$~(left),
   for $|\varsigma_{d,\ell}| \lesssim |\varsigma_u|\leq 2$ and $M_{H^\pm} = 80$, 200 and 500 GeV~(upper, middle and lower curves, respectively). The shaded horizontal bands denote the allowed experimental region at $1\sigma$~(dark green), $2\sigma$~(green), and $3\sigma$~(light green), respectively. The right panel shows the resulting upper bounds on $\varsigma_u$, as function of $M_{H^\pm}$.}
  \label{plot:C10}
\end{figure}

The dependence of $\bar R_{s\mu}$ on the alignment parameter $\varsigma_u$ with three typical charged-Higgs masses~(80, 200 and 500 GeV) is shown in Fig.~\ref{plot:C10}. One can see that, with the contributions from $C_S$ and $C_P$ ignored, the observable $\bar R_{s\mu}$ puts a strong constraint on the parameter $\varsigma_u$. For $M_{H^\pm}=80~(500)~\mathrm{GeV}$, a $95\%$ CL upper bound $|\varsigma_u|\leq 0.49~(0.97)$ is obtained, with the assumption $|\varsigma_{d,\ell}|\lesssim |\varsigma_u|$, which is stronger than the constraint from $R_b$~\cite{Jung:2010ik}. Since $C_{10}^{\mathrm{A2HDM}} \sim |\varsigma_u|^2$, this constraint is independent of any assumption about CP and, therefore, applies in the most general case.\footnote{Actually, the explicit correction factor given at the end of Eq.~\eqn{eq:BR_bar_CPcons} is valid only in the absence of new sources of CP violation beyond the SM. Taking the correct general relation into account, the upper-bounded parameter is $|\varsigma_u| \left\{\left[1 + y_s \cos{(2\phi_P-\phi_s^{\mathrm{NP}})} \right]/(1+y_s)\right\}^{1/4}\approx |\varsigma_u|$, where the phase $\phi_s^{\mathrm{NP}}$ denotes the CP-violating NP contribution to $B^0_s$--$\bar B^0_s$ mixing.} For larger charged-Higgs masses, the constraint becomes weaker as the NP effect starts to decouple, reflected by $\underset{x_{H^+}\to \infty}{\lim}C_{10}^{\mathrm{A2HDM}}=0$.

\subsubsection{Large $\boldsymbol{\varsigma_{d,\ell}}$}

When $\varsigma_{d}$ and $\varsigma_{\ell}$ are large, the scalar and pseudoscalar operators can induce a significant enhancement of the branching ratio. To see this explicitly, we vary $\varsigma_d$ and $\varsigma_\ell$ within the range $[-50, 50]$, and choose three representative values of $\varsigma_u$, $\varsigma_u=0,\pm1$. We also take three different representative sets of scalar masses:
\begin{eqnarray}
&\rm{Mass1}: &M_{H^\pm}= M_A = 80~\mathrm{GeV}, \quad M_H = 130~\mathrm{GeV}\,, \no \\
&\rm{Mass2}: &M_{H^\pm}= M_A = M_H = 200~\mathrm{GeV}\,, \no\\
&\rm{Mass3}: &M_{H^\pm}= M_A = M_H = 500~\mathrm{GeV}\,,
\end{eqnarray}
which cover the lower, intermediate, and upper range, respectively, of the allowed scalar spectrum.

\begin{figure}[t]
  \centering
  \includegraphics[width=\textwidth]{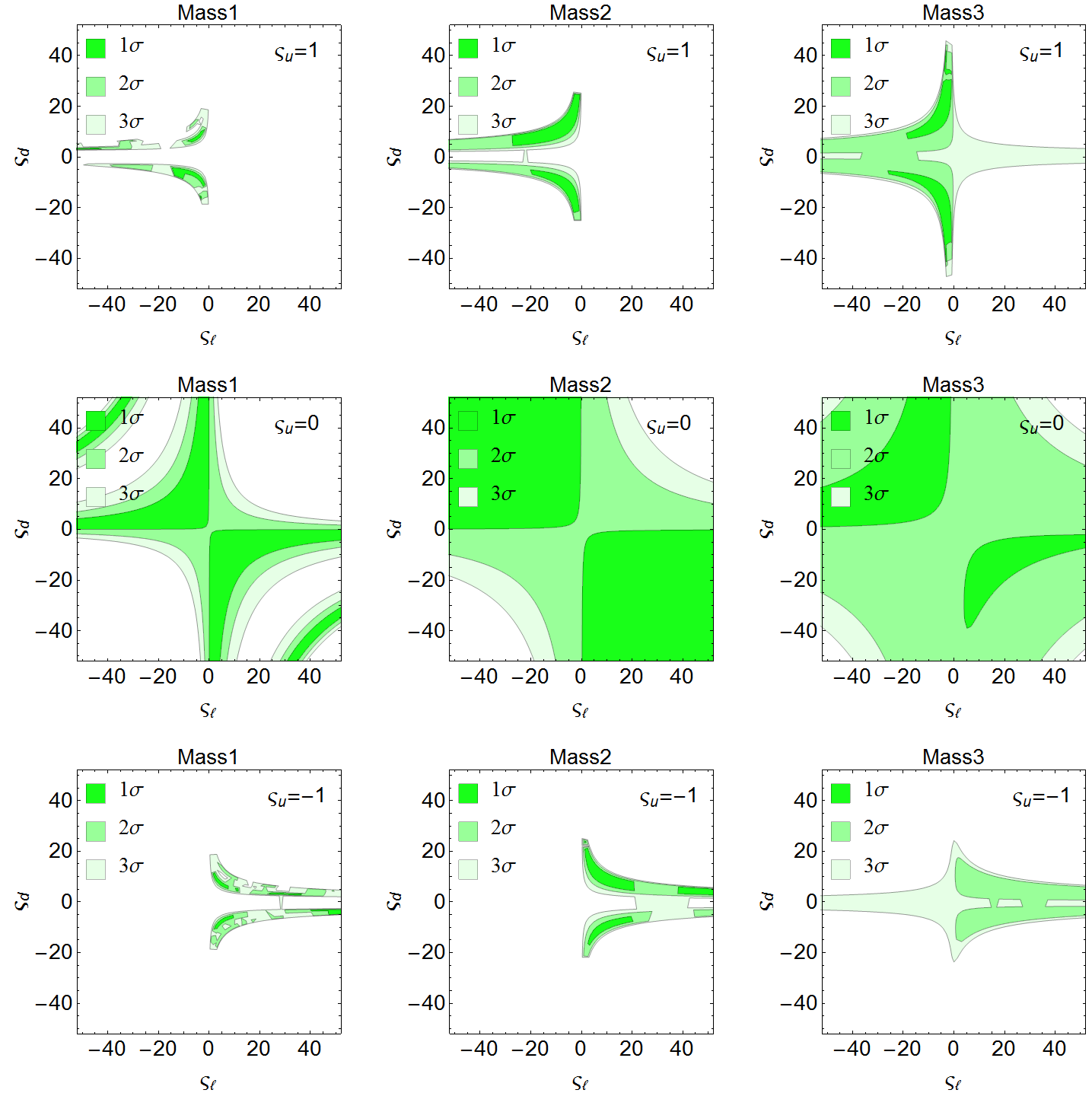}
  \caption{\small Allowed regions~(at $1\sigma$, $2\sigma$ and $3\sigma$) in the $\varsigma_d$--$\varsigma_\ell$ plane under the constraint from $\bar R_{s\mu}$, with three different assignments of the scalar masses and $\varsigma_u=0,\pm1$.}
  \label{plot:zdzl}
\end{figure}

With the above specification, we show in Fig.~\ref{plot:zdzl} the allowed regions in the $\varsigma_d$--$\varsigma_\ell$ plane under the constraint from $\bar R_{s\mu}$. One can see that, irrespective of the scalar masses, regions with large $\varsigma_d$ and $\varsigma_\ell$ are already excluded, especially when they have the same sign. The impact of $\varsigma_u$, even when varied within the small range $[-1,1]$, is found to be significant: a nonzero $\varsigma_u$ will exclude most of the regions allowed in the case with $\varsigma_u=0$, and changing the sign of $\varsigma_u$ will also flip that of $\varsigma_\ell$. This is mainly due to the factors $\varsigma_d^2\varsigma_u^{\ast}$ appearing in the functions $g_{2}^{(a)}$ and $g_{3}^{(a)}$ defined, respectively, by Eqs.~(\ref{eq:g2a_def}) and (\ref{eq:g3a_def}). It is also observed that the allowed regions expand with increasing scalar masses, as expected, since larger scalar masses make the NP contributions gradually decouple from the SM.

\subsection{$\mathcal{Z}_2$ symmetric models}
\label{sec:z2symmetrycase}

The five types of $\mathcal{Z}_2$-symmetric models listed in Table~\ref{tab:models} are particular cases of the CP-conserving A2HDM, with the three alignment factors $\varsigma_f$ reduced to a single parameter $\tan\beta=v_2/v_1\geq0$. In the particular scalar basis where the discrete $\mathcal{Z}_2$ symmetry is implemented, the scalar-potential couplings $\mu'_i$ and $\lambda'_i$ must be real, and $\mu'_3=\lambda'_6=\lambda'_7=0$; however, the rotation into the Higgs basis generates non-zero values of $\mu_3=-\frac{1}{2} \lambda_6 v^2$ and $\lambda_7$. Furthermore, the alignment condition is protected by the $\mathcal{Z}_2$ symmetry at any energy scale, which means that the misalignment parameter $C_R(M_W)$ does not contribute and the Higgs-penguin diagrams are free of divergences. Thus, for $\mathcal{Z}_2$-symmetric models, the ratio $\bar R_{s\mu}$ only involves seven free parameters: $M_{H^\pm}$, $M_H$, $M_A$, $\lambda_3$, $\lambda_7$, $\cos\tilde\alpha$, and $\tan\beta$.

A much more constrained case is the inert 2HDM, where the $\mathcal{Z}_2$ symmetry is imposed in the Higgs basis: all SM fields and $\Phi_1$ are even while $\Phi_2\to -\Phi_2$ under the $\mathcal{Z}_2$ transformation. This implies that there is no mixing between the CP-even neutral states $h$ and $H$, and the scalars $H$, $A$ and $H^{\pm}$ decouple from the fermions: $\cos\tilde\alpha=1$, $\lambda_6 = \lambda_7 =0$, $\varsigma_f=0$. Moreover, the couplings of $h$ to fermions and vector bosons are identical to the SM ones. Therefore, in the inert model $\bar R^{\rm inert}_{s\mu} = 1$.

For the other four types of $\mathcal{Z}_2$-symmetric models, we continue to use the assignments $\cos\tilde\alpha=0.95$ and $\lambda_3=\lambda_7=1$. One can easily check that the effects of $M_H$ and $M_A$ on $\bar R_{s\mu}$ are tiny, unless $\tan\beta$ is extremely small which is excluded by the flavour constraint $|\varsigma_u|\leq 2$. For simplicity, we fix them to be $M_{H}=M_{A}=500~\mathrm{GeV}$ in the following analysis.

\begin{figure}[t]
  \centering
  \includegraphics[width=\textwidth]{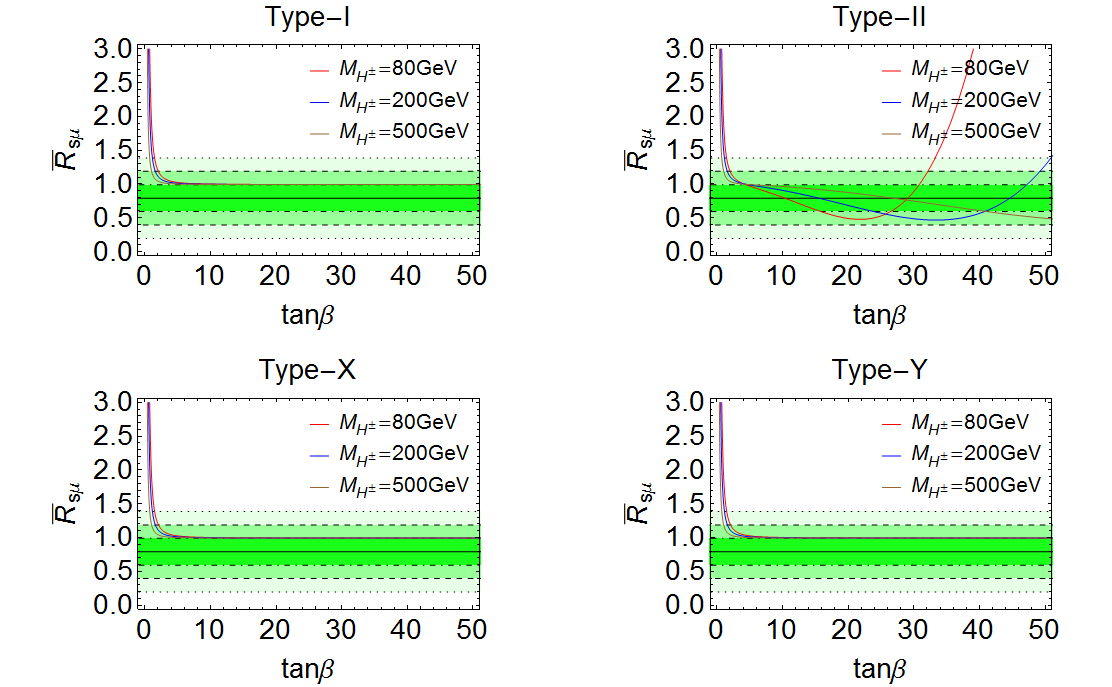}
  \caption{\small Dependence of $\bar R_{s\mu}$ on $\tan\beta$ for the 2HDMs of types I, II, X and Y. The upper, middle and lower curves correspond to $M_{H^\pm} = 80$, $200$ and $500~\mathrm{GeV}$, respectively. The horizontal bands denote the allowed experimental region at $1\sigma$~(dark green), $2\sigma$~(green), and $3\sigma$~(light green), respectively.}
  \label{plot:z2types}
\end{figure}

Fig.~\ref{plot:z2types} shows the dependence of $\bar R_{s\mu}$ on the parameter $\tan\beta$, for three representative values of the charged-Higgs mass: $M_{H^\pm}=80$, $200$ and $500~\mathrm{GeV}$. The four different panels correspond to the $\mathcal{Z}_2$-symmetric models of types I, II, X and Y, respectively. A lower bound $\tan\beta>1.6$ is obtained at $95\%$ CL under the constraint from the current experimental data on $\bar R_{s\mu}$. This implies $\varsigma_u = \cot{\beta} < 0.63$, which is stronger than the bounds obtained previously from other sources~\cite{Jung:2010ik,A2HDM:flavour}.

One can see that the predicted $\bar R_{s\mu}$ in the type-I, type-X and type-Y models are almost indistinguishable from each other and, in the large $\tan\beta$ region, approach the SM prediction, irrespective of the choices of scalar masses.
For the type-II model, on the other hand, an enhancement of $\bar R_{s\mu}$ is still possible in the large $\tan\beta$ region. This can be understood since the Wilson coefficients in the type-II model contain the factor $\tan^2\beta$ arising from the product of alignment parameters $\varsigma_f$, while in the other three models they contain at most one power of $\tan\beta$. So only the type-II model can receive a large $\tan\beta$ enhancement, which has been studied intensively in the literature~\cite{Logan:2000iv,Huang:2000sm,Bobeth:2001sq}. It is also interesting to note that in the type-II 2HDM with large $\tan\beta$ the $\Bqll$ branching ratios depend only on the charged-Higgs mass and $\tan\beta$~\cite{Logan:2000iv}.

\section{Conclusions}
\label{sec:conclusion}

In this paper, we have performed a detailed analysis of the rare decays $\Bqll$ within the general framework of the A2HDM. Firstly, we presented a complete one-loop calculation of the short-distance Wilson coefficients $C_{10}$, $C_S$ and $C_P$, which arise from various box and penguin diagrams, and made a detailed technical summary of our results and a comparison with previous calculations performed in particular limits or approximations. In order to make sure our results are gauge independent, the calculations were carried out in both the Feynman and the unitary gauges.

With the current data on $\overline{\mathcal{B}}(\Bsmm)$ taken into account, we have also investigated the impact of various model parameters on the branching ratios and studied the phenomenological constraints imposed by present data. The resulting information about the model parameters will be crucial for the model building and is complementary to the collider searches for new scalar resonances in the near future.

When $|\varsigma_{d,\ell}| \lesssim |\varsigma_u|$, the contributions to $\overline{\mathcal{B}}(\Bsmm)$ from the scalar and pseudoscalar operators are negligible compared to the leading Wilson coefficient $C_{10}$. Since $C_{10}^{\mathrm{A2HDM}}\sim |\varsigma_u|^2$, the measured $\overline{\mathcal{B}}(\Bsmm)$ branching ratio implies then an upper bound on the up-family alignment parameter, which only depends on the charged Higgs mass. At 95\% CL, we obtain:
\begin{equation}
|\varsigma_u| \,\leq\, 0.49 \;\: (0.97)\, , \qquad\qquad \mathrm{for} \quad M_{H^\pm} = 80~(500)~\mathrm{GeV} \quad \mathrm{and} \quad\, |\varsigma_{d,\ell}| \lesssim |\varsigma_u|\,.
\end{equation}
This bound is stronger than the constraints obtained previously from other sources~\cite{Jung:2010ik,A2HDM:flavour}.

The role of the scalar and pseudoscalar operators becomes much more important for large values of $|\varsigma_{d,\ell}|$. This region of parameter space was previously explored within the context of the type-II 2HDM, where these contributions are enhanced by a factor $\tan^2{\beta}$. Our analysis agrees with previous results in the type-II case and shows, moreover, that this $\tan^2{\beta}$ enhancement is absent in the $\mathcal{Z}_2$-symmetric models of types I, X and Y, which approach the SM prediction for large values of $\tan{\beta}$. From the current experimental data on $\bar R_{s\mu}$, we derive
the 95\% CL bound:
\begin{equation}
\tan\beta\; >\; 1.6\, , \qquad\qquad \mathrm{for \ 2HDMs \ of \ types \ I, \ II, \ X \ and \ Y.}
\end{equation}
This implies $\varsigma_u = \cot{\beta} < 0.63$, which is also stronger than the bounds obtained previously from other sources~\cite{Jung:2010ik,A2HDM:flavour}.

The enhancement of the scalar and pseudoscalar contributions at large values of
$|\varsigma_{d,\ell}|$ is present in the most general A2HDM scenario and could give rise to interesting phenomenological signals. To exemplify this possibility, we have analyzed the ratio $\bar R_{s\mu}$ in the simpler CP-conserving case, showing the important impact of the A2HDM corrections whenever enhanced Yukawa couplings to leptons and down-type quarks are present. The resulting constraints on the alignment parameters are given in Fig.~\ref{plot:zdzl}.

It would be interesting to analyze the possible impact of the new CP-violating phases present within the A2HDM framework, at large values of $|\varsigma_{d,\ell}|$. They could generate sizeable phases $\phi_P$ and $\phi_S$ in Eqs.~\eqn{eq:P} and \eqn{eq:S}, which could manifest themselves in the time-dependence of the $\Bsmm$ decay amplitude~\cite{Buras:2013uqa}. To quantify the possible size of this effect requires a more careful assessment of the allowed parameter space of the A2HDM, which we plan to further investigate in future works.

\section*{Acknowledgements}
We are grateful to Alejandro Celis and Victor Ilisie for useful discussions on the 2HDM parameters.
This work was supported in part by the National Natural Science Foundation of China~(NSFC) under contract No.~11005032, the Spanish Government and ERDF funds from the EU Commission~[Grants FPA2011-23778 and CSD2007-00042~(Consolider Project CPAN)] and by Generalitat Valenciana under Grant No. PROMETEOII/2013/007. X.~Q. Li was also supported by the Specialized Research Fund for the Doctoral Program of Higher Education of China~(Grant No.~20104104120001) and by the Scientific Research Foundation for the Returned Overseas Chinese Scholars, State Education Ministry.

\begin{appendix}

\section*{Appendix: Scalar-penguin results within the A2HDM}
\label{app:input}

The coefficients $\hat{C}^{\varphi^0_i}$, defined in Eq.~\eqn{eq:ScalarPenguinWCs}, are given by
\be
\hat{C}^{\varphi^0_i}\; = \; \frac{m_t^2}{M_{\varphi^0_i}^2}\;\left\{\,
\sum_{k=1}^{18}\; C^{k,\,\varphi^0_i} \; + \; \frac{1}{2}\,
(\varsigma_u-\varsigma_d)\,(1+\varsigma^*_u\varsigma_d)\, (\cR_{i2} + i \cR_{i3})\; \cC \,\right\}\, ,
\ee
where the last term is the tree-level contribution from the local operator in Eq.~\eqn{eq:FCNCop}. We detail next the contributions $C^{k,\,\varphi^0_i}$ from the separate diagrams~($k=1,\cdots,18$) shown in Fig.~\ref{fig:hiA2HDM}.

The gauge-independent coefficients are:
\begin{align}
C^{1,\,\varphi^0_i} &\; =\; \frac{y^{\varphi^0_i}_u}{4}\, \Bigg\{\varsigma_d\,\varsigma_u^*\,
 \frac{x_t}{x_{H^+}-x_t}\, \left[1-\frac{x_{H^+}}{x_{H^+}-x_t}(\ln x_{H^+}-\ln x_t)\right]\nonumber\\[0.2cm]
 & \hspace{1.5cm} + |\varsigma_u|^2\,\frac{x_t}{2(x_{H^+}-x_t)^2}\,\left[\frac{3 x_t- x_{H^+}}{2}+\frac{x_{H^+}(x_{H^+}-2 x_t)}{x_{H^+}-x_t}(\ln x_{H^+}-\ln x_t)\right]\Bigg\}\nonumber\\[0.2cm]
 &\; + \frac{y^{\varphi^0_i*}_u}{4}\,\Bigg\{\varsigma_d\,\varsigma_u^*\, \left[\Lambda+\frac{x_t}{x_{H^+}-x_t}-\frac{x_{H^+}^2}{(x_{H^+}-x_t)^2}\ln x_{H^+}+\frac{x_t(2 x_{H^+}-x_t)}{(x_{H^+}-x_t)^2}\ln x_t\right]\nonumber\\[0.2cm]
 & \hspace{1.5cm} + |\varsigma_u|^2\,\frac{x_t}{2(x_{H^+}-x_t)^2}\,\left[\frac{3 x_{H^+}-x_t}{2}-\frac{x_{H^+}^2}{x_{H^+}-x_t}(\ln x_{H^+}-\ln x_t)\right]\Bigg\}\,,\\[0.3cm]
C^{2,\,\varphi^0_i} &\; =\; \frac{s^2_W\lambda^{\varphi^0_i}_{H^+H^-}}{4\pi\alpha(x_{H^+}-x_t)}\,
 \Bigg\{\varsigma_d\,\varsigma_u^*\,\left[\frac{x_t}{x_{H^+}-x_t}(\ln x_{H^+}-\ln x_t) -1 \right] \nonumber\\[0.2cm]
 & \hspace{3.4cm} + |\varsigma_u|^2\,\left[\frac{x_t^2}{2(x_{H^+}-x_t)^2}(\ln x_{H^+}-\ln x_t) + \frac{x_{H^+}-3 x_t}{4(x_{H^+}-x_t)}\right]\Bigg\}\,,\\[0.3cm]
C^{3,\,\varphi^0_i} &\; =\; \frac{y_d^{\varphi^0_i}}{4}\,\varsigma_d\,\varsigma_u^*\,\left[-\Lambda
 + \frac{x_{H^+}}{x_{H^+}-x_t}\ln x_{H^+} - \frac{x_t}{x_{H^+}-x_t}\ln x_t\right]\,,\\[0.3cm]
C^{4,\,\varphi^0_i}&\; =\;
C^{7, \,\varphi^0_i} \; =\;
C^{8, \,\varphi^0_i}
\; =\; 0 \,.
\end{align}

In the unitary gauge, we find:
\begin{align}
C_{\rm Unitary}^{5,\, \varphi^0_i} &\; =\; \frac{1}{4}\, \Bigg\{y^{\varphi^0_i *}_u\,\left[\Lambda -\frac{5 x_t^2-13
 x_t+2}{4(x_t-1)^2}-\frac{2 x_t^3-6 x_t^2+9 x_t-2}{2(x_t - 1)^3}\ln x_t\right]\nonumber\\[0.2cm]
 & \hspace{1.1cm} + y^{\varphi^0_i}_u\,\left[\frac{\Lambda}{2}-\frac{2 x_t^2-x_t-7}{4(x_t-1)^2}-\frac{x_t^3-3 x_t^2+3 x_t+2}{2(x_t -1)^2}\ln x_t\right]\Bigg\}\,,\\[0.3cm]
C_{\rm Unitary}^{6,\, \varphi^0_i} &\; =\; \frac{\lambda^{\varphi^0_i}_{W^+W^-}}{8}\, \left[-3 \Lambda +
 \frac{x_t^2-2 x_t-11}{2(x_t-1)^2}+\frac{3 x_t (x_t^2-3 x_t+4)}{(x_t - 1)^3}\ln x_t \right]\,,\\[0.3cm]
C_{\rm Unitary}^{9,\, \varphi^0_i} &\; =\; \frac{\lambda^{\varphi^0_i}_{H^+W^-}}{8}\,
 \varsigma_u^*\,\left[\frac{1}{2}-\Lambda+\frac{x_{H^+}(x_{H^+}+2)\ln x_{H^+}}{(x_{H^+}-1)(x_{H^+}-x_t)}-\frac{x_t(x_t+2)\ln x_t}{(x_t-1)(x_{H^+}-x_t)}\right]\,,\\[0.3cm]
C_{\rm Unitary}^{10,\, \varphi^0_i} &\; = \; \frac{\lambda^{\varphi^0_i *}_{H^+W^-}}{4}\,\Bigg\{\varsigma_d
 \left[-\Lambda + \frac{x_{H^+}\ln x_{H^+}}{x_{H^+}-x_t} - \frac{x_t\ln x_t}{x_{H^+}-x_t}\right] - \frac{\varsigma_u}{2}\left[\frac{x_t(x_{H^+}x_t - 4 x_{H^+}+3 x_t)}{(x_t-1)(x_{H^+}-x_t)^2}\ln x_t \right. \nonumber\\[0.2cm]
 & \hspace{3.2cm} \left. + \frac{x_{H^+}}{x_{H^+}- x_t}-\frac{x_{H^+}(x_{H^+}x_t-3x_{H^+}+2 x_t)}{(x_{H^+}-1)(x_{H^+}-x_t)^2}\ln x_{H^+}\right]\Bigg\}\,.
\end{align}

In the Feynman gauge the results are:
\begin{align}
C_{\rm Feynman}^{5,\,\varphi^0_i} &\; =\; \frac{1}{8(x_t-1)^2}\,\Bigg\{y^{\varphi^0_i *}_u\,\left[3 x_t - 1
 + \frac{2(1-2 x_t)\ln x_t}{x_t-1}\right] + y^{\varphi^0_i}_u\,\left[3 - x_t - \frac{2\ln x_t}{x_t-1}\right] \Bigg\}\,,\\[0.3cm]
C_{\rm Feynman}^{6,\,\varphi^0_i} &\; =\; \frac{\lambda^{\varphi^0_i}_{W^+W^-}}{4(x_t-1)^2}\,\left[ \frac{2
 x_t}{x_t-1}\ln x_t - x_t -1 \right]\,,\\[0.3cm]
C_{\rm Feynman}^{9,\,\varphi^0_i} &\; = \; \frac{\varsigma_u^*\,\lambda^{\varphi^0_i}_{H^+W^-}}{8(x_{H^+}-x_t)}\,
 \left[\frac{x_{H^+}-x_t}{(x_{H^+}-1)(x_t - 1)} + \frac{x_{H^+}(3x_{H^+} - 2)\ln x_{H^+}}{(x_{H^+}-1)^2} - \frac{x_t(3 x_t-2)\ln x_t}{(x_t - 1)^2}\right]\,,\\
C_{\rm Feynman}^{10,\,\varphi^0_i} &\; =\; \frac{\lambda^{\varphi^0_i*}_{H^+W^-}}{4(x_{H^+}-x_t)}\,\Bigg\{\varsigma_d\,
 \left[\frac{x_t\ln x_t}{x_t-1}-\frac{x_{H^+}\ln x_{H^+}}{x_{H^+}-1}\right] + \frac{\varsigma_u}{2}\,\left[\frac{x_{H^+}}{x_{H^+}-1}+\frac{x_t(4 x_{H^+}-3 x_t)\ln x_t}{(x_t-1)(x_{H^+}-x_t)}\right. \nonumber\\[0.2cm]
 & \hspace{3.5cm} \left. - \frac{x_{H^+}(4 x_{H^+}^2-3 x_{H^+}x_t-3 x_{H^+}+2 x_t)}{(x_{H^+}-1)^2(x_{H^+}-x_t)}\ln x_{H^+}\right] \Bigg\}\,,\\[0.3cm]
C_{\rm Feynman}^{11,\,\varphi^0_i} & \; =\;  \frac{1}{4}\,\Bigg\{y^{\varphi^0_i *}_u\,\left[\Lambda-\frac{x_t(5
 x_t-7)}{4(x_t-1)^2}-\frac{x_t(2 x_t^2-6 x_t+5)}{2(x_t-1)^3}\ln x_t\right]\nonumber\\[0.2cm]
 & \hspace{1.0cm} - \frac{y^{\varphi^0_i}_u}{2}\,\left[\frac{x_t(x_t-3)}{2(x_t-1)^2}+\frac{x_t}{(x_t-1)^3}\ln x_t\right]\Bigg\}\,,\\[0.3cm]
C_{\rm Feynman}^{12,\,\varphi^0_i} &\; =\; \frac{s^2_W\lambda^{\varphi^0_i}_{G^+G^-}}{16\pi\alpha(x_t-1)^2}\,
 \left[x_t-3 - \frac{2x_t(x_t-2)}{x_t-1}\ln x_t\right]\,,\\[0.3cm]
C_{\rm Feynman}^{13,\,\varphi^0_i} &\; = \; \frac{y^{\varphi^0_i}_d}{4}\,\left[-\Lambda + \frac{x_t}{x_t-1}\ln
 x_t\right]\,,\\[0.3cm]
C_{\rm Feynman}^{14,\,\varphi^0_i} &\; =\; 0 \,,\\[0.3cm]
C_{\rm Feynman}^{15,\,\varphi^0_i} &\; = \; \frac{s^2_W\varsigma_u^*\,\lambda^{\varphi^0_i}_{H^+G^-}}{8
\pi\alpha (x_{H^+}-x_t)}\,\left[\frac{x_{H^+}-x_t}{(x_{H^+}-1)(x_t-1)} + \frac{x_t(x_t-2)}{(x_t-1)^2}\ln x_t - \frac{x_{H^+}(x_{H^+}-2)}{(x_{H^+}-1)^2}\ln x_{H^+}\right]\,,\\
C_{\rm Feynman}^{16,\,\varphi^0_i} &\; = \; \frac{s^2_W\lambda^{\varphi^0_i*}_{H^+G^-}}
 {8\pi\alpha(x_{H^+}-x_t)}\, \Bigg\{2\,\varsigma_d\left[\frac{x_t}{x_t-1}\ln x_t - \frac{x_{H^+}}{x_{H^+}-1}\ln x_{H^+}\right] + \varsigma_u\left[\frac{x_t^2\ln x_t}{(x_t-1)(x_{H^+}-x_t)} \right. \nonumber \\[0.2cm]
 & \hspace{4.0cm} \left. +\frac{x_{H^+}}{x_{H^+}-1} - \frac{x_{H^+}(x_{H^+}x_t + x_{H^+}-2 x_t)}{(x_{H^+}-1)^2(x_{H^+}-x_t)}\ln x_{H^+}\right] \Bigg\}\,, \\[0.3cm]
C_{\rm Feynman}^{17,\,\varphi^0_i} &\; = \; \frac{\lambda^{\varphi^0_i}_{G^+W^-}}{8(x_t-1)^2}\,\left[\frac{5-7
 x_t}{2} + \frac{x_t(3x_t-2)}{x_t-1}\ln x_t \right]\,,\\[0.3cm]
C_{\rm Feynman}^{18,\,\varphi^0_i} &\; = \; \frac{\lambda^{\varphi^0_i}_{G^+W^-}}{8(x_t-1)^2}\,\left[\frac{9
 x_t-11}{2} - \frac{x_t(5x_t-6)}{x_t-1}\ln x_t \right]\,.
\end{align}
Here $\Lambda =-\frac{2\,\mu^{D-4}}{D-4} -\gamma_E + \ln{(4\pi)} - \ln{\left(\frac{M_W^2}{\mu^2}\right)} +1$, and the (rescaled) cubic coupling constants are defined, respectively, as
\begin{align}
\lambda_{W^{+}W^{-}}^{\varphi^0_i} &\; =\; \lambda_{G^{+}W^{-}}^{\varphi^0_i}\; =\;\cR_{i1}\,,
\\[0.2cm]
\lambda_{H^{+}W^{-}}^{\varphi^0_i} &\; =\;  \cR_{i2}-i\cR_{i3}\,,\\[0.2cm]
\lambda_{H^{+}H^{-}}^{\varphi^0_i} &\; =\; 	  \lambda_{3}\,\cR_{i1}+\lambda^R_7\,\cR_{i2}
 -\lambda^I_7\,\cR_{i3}\,,\\[0.2cm]
\lambda_{G^{+}G^{-}}^{\varphi^0_i} &\; =\; 2\lambda_{1}\,\cR_{i1}+ \lambda^R_6 \,\cR_{i2}- \lambda^I_6\,\cR_{i3}\; =\; \frac{M^2_{\varphi_i^0}}{v^2}\,\cR_{i1}\,,
\\[0.2cm]
\lambda_{H^{+}G^{-}}^{\varphi^0_i} &\; =\;  \lambda_{6}\,\cR_{i1}+\frac{1}{2}\left(\lambda_{4}+2\lambda_{5}\right)\cR_{i2}
 -\frac{i}{2}\left(\lambda_{4}-2\lambda_{5}\right)\cR_{i3}
\; = \;\frac{M^2_{\varphi_i^0}-M^2_{H^+}}{v^2}\,\left(\cR_{i2}-i\cR_{i3}\right)
 \,.
\end{align}

The functions $g_0^{\phantom{()}}(x_t,x_{H^+},\varsigma_u,\varsigma_d)$ and $g_{j}^{(a)}(x_t,x_{H^+},\varsigma_u,\varsigma_d)$ introduced in Eq.~(\ref{eq:scalarPenguinRes}) are gauge independent. For $g_0^{\phantom{()}}(x_t,x_{H^+},\varsigma_u,\varsigma_d)$ we find
\begin{equation}
g_0^{\phantom{()}}(x_t,x_{H^+},\varsigma_u,\varsigma_d) \;=\; \frac{\pi\alpha\, C^{2,\,\varphi^0_i}}{s^2_W\,\lambda_{H^{+}H^{-}}^{\varphi^0_i}}\,,
\end{equation}
while the functions $g_{j}^{(a)}(x_t,x_{H^+},\varsigma_u,\varsigma_d)$ are given, respectively, as:
\begin{eqnarray}
g_{1}^{(a)}(x_t,x_{H^+},\varsigma_u,\varsigma_d) &=&
   -\frac{3}{4}+\varsigma_d\,\varsigma^*_u\, \frac{x_t}{x_{H^+}-x_t}\left[1
   -\frac{x_{H^+}}{x_{H^+}-x_t}(\ln x_{H^+}-\ln x_t)\right] \no \\[0.2cm]
&&\hspace{-1cm}
   +|\varsigma_u|^2\, \frac{x_t}{2(x_{H^+}-x_t)^2}\left[\frac{x_{H^+}+x_t}{2} - \frac{x_{H^+}x_t}{x_{H^+}-x_t}(\ln x_{H^+}-\ln x_t)\right] \,,\label{eq:g1a_def}\\[0.3cm]
g_{2}^{(a)}(x_t,x_{H^+},\varsigma_u,\varsigma_d) &=&
   \varsigma_d^2 \varsigma_u^*  f_1(x_t,x_{H^+})
   +\varsigma_d(\varsigma_u^*)^2 f_2(x_t,x_{H^+} ) \no \\[0.2cm]
 &&\hspace{-1cm}
   +\varsigma_d |\varsigma_u|^2 f_3(x_t,x_{H^+} )
   +\varsigma_u |\varsigma_u|^2 f_4(x_t,x_{H^+} )
   -\varsigma^*_u |\varsigma_u|^2 f_5(x_t,x_{H^+} )
\no \\[0.2cm]
 &&\hspace{-1cm}
   +\varsigma_u f_6(x_t,x_{H^+} )
   -\varsigma_u^* f_7(x_t,x_{H^+} )
   +\varsigma_d f_1(x_t,x_{H^+} )\,,\label{eq:g2a_def}\\[0.3cm]
g_{3}^{(a)}(x_t,x_{H^+},\varsigma_u,\varsigma_d) &=&
   \varsigma_d^2 \varsigma_u^*  f_1(x_t,x_{H^+})
   -\varsigma_d(\varsigma_u^*)^2 f_2(x_t,x_{H^+} ) \no \\[0.2cm]
 &&\hspace{-1cm}
   +\varsigma_d |\varsigma_u|^2 f_3(x_t,x_{H^+} )
   +\varsigma_u |\varsigma_u|^2 f_4(x_t,x_{H^+} )
   +\varsigma^*_u |\varsigma_u|^2 f_5(x_t,x_{H^+} )
\no \\[0.2cm]
 &&\hspace{-1cm}
   +\varsigma_u f_6(x_t,x_{H^+} )
   +\varsigma_u^* f_7(x_t,x_{H^+} )
   +\varsigma_d f_1(x_t,x_{H^+} )\,. \label{eq:g3a_def}
\end{eqnarray}
The functions $g_{j}^{(b)}(x_t,x_{H^+},\varsigma_u,\varsigma_d)$ are zero in the unitary gauge, because they are all related to Goldstone boson vertices.
In the Feynman gauge, they are given, respectively, as
\begin{eqnarray}
g_{1,\rm Feynman}^{(b)}(x_t,x_{H^+},\varsigma_u,\varsigma_d) &=&
 \frac{1}{8(x_t-1)^2}\,\left[\frac{
 x_t-3}{2} - \frac{x_t(x_t-2)} {x_t-1} \ln x_t \right]\,,\\[0.2cm]
g_{2,\rm Feynman}^{(b)}(x_t,x_{H^+},\varsigma_u,\varsigma_d) &=&
 \varsigma_d f_8(x_t,x_{H^+})+\varsigma_u\, f_9(x_t,x_{H^+})+
 \varsigma^*_u\, f_{10}(x_t,x_{H^+}) \,,\\[0.2cm]
g_{3,\rm Feynman}^{(b)}(x_t,x_{H^+},\varsigma_u,\varsigma_d) &=&
 \varsigma_d f_8(x_t,x_{H^+})+\varsigma_u\, f_9(x_t,x_{H^+}) - \varsigma^*_u\, f_{10}(x_t,x_{H^+}) \,.
\end{eqnarray}
Here the functions $f_j(x_t,x_{H^+})$ are defined, respectively, as
\begin{align}
 f_1(x_t,x_{H^+} )&= \frac{1}{2 (x_{H^+}- x_t)}\left[-x_{H^+}+x_t +x_{H^+} \ln x_{H^+}-x_t \ln x_t\right]\,, \\[0.2cm]
 f_2(x_t,x_{H^+}) &= \frac{1}{2(x_{H^+}-x_t)} \left[x_t - \frac{x_{H^+} x_t}{x_{H^+}-x_t}(\ln x_{H^+}-\ln x_t)\right]\,, \\[0.2cm]
 f_3(x_t,x_{H^+}) &=\frac{1}{2(x_{H^+}-x_t)}
      \left[x_{H^+}-\frac{x_{H^+}^2 \ln x_{H^+}}{x_{H^+}-x_t}+\frac{x_t (2x_{H^+}- x_t) \ln x_t}{x_{H^+}-x_t}\right]\,, \\[0.2cm]
 f_4(x_t,x_{H^+}) &=\frac{1}{4(x_{H^+}-x_t)^2}
     \left[\frac{x_t \left(3 x_{H^+}-x_t\right)}{2}-\frac{x_{H^+}^2 x_t}{x_{H^+}- x_t}(\ln x_{H^+}-\ln   x_t)\right]\,, \\[0.2cm]
 f_5(x_t,x_{H^+}) &=\frac{1}{4(x_{H^+}-x_t)^2}
   \left[\frac{x_t (x_{H^+}-3 x_t)}{2}-\frac{x_{H^+} x_t (x_{H^+}-2 x_t)}{x_{H^+}- x_t}(\ln x_{H^+}-\ln   x_t)\right]\,, \\[0.2cm]
 f_6(x_t,x_{H^+}) &=\frac{1}{2(x_{H^+}-x_t)} \left[\frac{x_t\left(x_t^2-3x_{H^+}x_t+9x_{H^+}-5 x_t-2\right)}{4
   (x_t-1)^2}\right.\no \\[0.2cm]
  & \left. +\frac{x_{H^+} \left(x_{H^+} x_t-3x_{H^+}+2
   x_t\right) }{2 (x_{H^+}-1)(x_{H^+}- x_t)}\ln x_{H^+}\right.\no \\[0.2cm]
  &\hspace{-1.2cm}\left.+\frac{x_{H^+}^2 \left(-2 x_t^3+6 x_t^2-9
   x_t+2\right)+3 x_{H^+} x_t^2 (x_t^2-2 x_t+3)-x_t^2
   \left(2 x_t^3-3 x_t^2+3 x_t+1\right)}{2 (x_t-1)^3
   (x_{H^+}- x_t)}\ln x_t\right]\,, \\[0.2cm]
 f_7(x_t,x_{H^+}) &=\frac{1}{2(x_{H^+}-x_t)}
   \left[\frac{\left(x_t^2+x_t-8\right) (x_{H^+}- x_t)}{4 (x_t-1)^2}-\frac{x_{H^+}   (x_{H^+}+2) }{2 (x_{H^+}-1)}\ln x_{H^+}\right.\no \\[0.2cm]
   & \left.+\frac{x_{H^+} \left(x_t^3-3 x_t^2+3 x_t+2\right)+3 \left(x_t-2\right) x_t^2}{2 (x_t-1)^3}\ln x_t\right]\,, \\[0.2cm]
  f_8(x_t,x_{H^+}) &= \frac{1}{4(x_{H^+}-x_t)} \left[\frac{x_t\ln x_t}{x_t-1}-\frac{x_{H^+}\ln x_{H^+}}{x_{H^+}-1}\right]\,,\no \\[0.2cm]
 f_9(x_t,x_{H^+}) &= \frac{1}{8(x_{H^+}-x_t)}\left[\frac{x_{H^+}}{x_{H^+}-1} + \frac{x_t^2\ln x_t}{(x_t-1)(x_{H^+}-x_t)} - \frac{x_{H^+}(x_{H^+}x_t + x_{H^+}-2 x_t)}{(x_{H^+}-1)^2(x_{H^+}-x_t)}\ln x_{H^+}\right]\,, \\[0.2cm]
 f_{10}(x_t,x_{H^+}) &= \frac{1}{8(x_{H^+}-x_t)}\left[\frac{x_{H^+}-x_t}{(x_{H^+}-1)(x_t-1)} + \frac{x_t(x_t-2)}{(x_t-1)^2}\ln x_t - \frac{x_{H^+}(x_{H^+}-2)}{(x_{H^+}-1)^2}\ln x_{H^+}\right]\,.
\end{align}

\end{appendix}

\end{document}